\documentclass[twocolumn, superscriptaddress, pra, amsmath,amssymb, aps, floatfix, 10pt, nofootinbib]{revtex4-2}

\usepackage{graphicx}
\usepackage{svg}
\usepackage{dcolumn}
\usepackage{bm}
\usepackage{dutchcal}
\usepackage[protrusion=true, expansion=true]{microtype}

\usepackage{braket}
\usepackage{bm,xcolor}
\usepackage{appendix}
\usepackage{amsmath}
\usepackage{amsfonts}
\usepackage{amssymb}
\usepackage{extarrows}
\usepackage{xfrac}
\usepackage{mathrsfs}
\usepackage[draft]{changes}

\usepackage{verbatim}
\usepackage{changes}

\usepackage{siunitx}
\definecolor{navyblue}{rgb}{0.0, 0.0, 0.5}
\usepackage[colorlinks=true, breaklinks=false, linkcolor=navyblue, urlcolor=navyblue, citecolor=navyblue]{hyperref}

\begin{document}

\title{Leakage Mobility and Passive Leakage Removal in Transmons with Tunable Couplers}

\author{Taneli Tolppanen}
\thanks{These authors contributed equally to this work.}
\affiliation{Nano and Molecular Systems Research Unit, University of Oulu, FI-90014 Oulu, Finland}
\author{Gonzalo Martín-Vázquez}
\thanks{These authors contributed equally to this work.}
\affiliation{Nano and Molecular Systems Research Unit, University of Oulu, FI-90014 Oulu, Finland}
\affiliation{Departamento de Física Aplicada II, Universidad de Sevilla, E-41012 Sevilla, Spain}
\author{Sasu Tuohino}
\affiliation{QCD Labs, QTF Centre of Excellence, Department of Applied Physics, Aalto University, P.O. Box 13500, FI-00076 Aalto, Finland}
\affiliation{Nano and Molecular Systems Research Unit, University of Oulu, FI-90014 Oulu, Finland}
\author{Matti Silveri}
\affiliation{Nano and Molecular Systems Research Unit, University of Oulu, FI-90014 Oulu, Finland}

\date{\today}

\begin{abstract}
Qubit leakage is a noticeable source of errors for quantum computing. In quantum processors,  leakage excitations traveling between qubits generate correlated errors and perturb gate implementations. Leakage mobility can also be utilized for creating dedicated leakage removal pathways and removal units. To quantitatively characterize leakage mobility and to guide better design of processor architectures, we study here leakage dynamics in transmons with tunable couplers through numerical and analytical methods. Even if the couplers are tuned to cancel the single-excitation exchange or the $\textrm{ZZ}$ interaction, the leakage hopping rates still persists in the range of~\SIrange{0.8}{10}{\mega\hertz} due to transmon nonlinearity. In typical operation regimes, however, transmon frequency detuning localizes leakage excitations. The next-nearest-neighbor transmons can be still be near-resonant opening leakage tunneling channels. To suppress longer-range hopping, we find that the frequency spread of the next-nearest-neighbor transmons needs to be in the range of~\SIrange{1}{4}{\mega\hertz}. Utilizing leakage mobility, we propose two passive leakage removal units. One is based on a tunable coupler and a pumped transmon, and another on a junction readout scheme. Based on realistic experimental parameters, our results on selectively mobilizing or localizing leakage excitations are readily applicable in superconducting quantum devices. 
\end{abstract}

\maketitle

\section{Introduction}
Qubit population leakage beyond the qubit subspace is a considerable error source for practical quantum computing. It is a challenge even for quantum error correction~\cite{miao_overcoming_2023} since typical error correction codes are protected only against the qubit errors~\cite{Terhal15}. Often leakage is viewed only from the perspective of minimizing its creation sources~\cite{Motzoi09, Hyyppa24} and maximizing removal~\cite{varbanov20, miao_overcoming_2023, Marques23, Marxer26}. However, in multi-qubit systems, such as quantum processors, leakage mobility is also an important phenomena. Leakage excitations traveling among qubits can cause correlated errors~\cite{Fowler13}, induce leakage excitation multiplication~\cite{varbanov20}, decrease gate fidelity~\cite{gambetta_2011, chen_2016, rol_2019} and increase general systemic randomness. Despite apparent negative implications, leakage mobility can also be utilized for realizing leakage transport away from critical qubits or for creating leakage removal units via natural dynamic switching between data and auxiliary qubits~\cite{Camps24}. 

Leakage mobility depends on the qubit lattice architecture, couplers, and operational parameters. Our focus is in passive leakage mobility in contrast to transfer or generation of leakage during one- or two-qubit gates~\cite{varbanov20}. Gates instances are seen here as events altering leakage populations or changing mobility momentarily.  In other words, our focus is in the question how leakage excitations travel during qubit idling when couplers are set to a parameter position where qubit-qubit interactions are as minimal as possible. We view that the coupler OFF positioning as the fundamental operational setting describing the passive leakage mobility phenomena of a processor. 

The overall aim is to quantitatively characterize leakage mobility and to guide better design of quantum processor. To this end, we study superconducting transmons with tunable couplers~\cite{chen_2014, yan_2018, li_2020, sung_2021, marxer_2023} as the most popular superconducting circuit based architecture~\cite{Kjaergaard20}. The results and methods are applicable to other comparable systems, such as to coupled fluxonium qubits~\cite{Ding23, Zhang24, Rosenfeld24}. We combine numerical methods to analytical Schrieffer-Wolf perturbation theory.

Tunable couplers realize their OFF-positions by canceling effective single-excitation exchange or $\textrm{ZZ}$ interactions via destructive interference between multiple different physical interaction pathways. Interference is sensitive to exact values of phases and strength of interactions. Transmon nonlinearity implies that these phases and matrix elements have different values inside the qubit subspace and beyond it. A central result is then that even if couplers are tuned to cancel the single-excitation exchange or the $\textrm{ZZ}$ interaction, the leakage hopping rates still persists in the range of~\SIrange{0.8}{10}{\mega\hertz}. The only practical method then to suppress leakage mobility is to have frequency detuning between the leakage levels, which is utilized in current standard processors~\cite{mundada_2019, Marxer26, wesdorp26}. 

Even if neighboring transmons are sufficiently off-resonant, next-nearest transmons can still be near-resonant. Relevant situation is the so called ABA frequency architecture~\cite{Karamlou24,Marxer26, wesdorp26}. In current superconducting quantum processors, each transmon is coupled to \SIrange{2}{4}{} other transmons so that the number of next-nearest sites can be as high as \SIrange{3}{8}{} amplifying the need for detailed leakage tunneling control. We find that the leakage excitation tunneling rates to next-nearest transmons are of the order~\SI{0.1}{\mega\hertz}. Thus, the scale of frequency perturbations needed to create localized leakage excitations is of the order~\SI{1}{\mega\hertz}. 

On the flip side, mobile leakage excitations in the presence of canceled qubit subspace interactions can be utilized for creating effective leakage removal units. We show this for two setups. The first leakage removal unit is based on combination of a pumped and dissipated transmon with a tunable coupler. The second proposed leakage removal unit other derives its structure from a junction readout setup~\cite{chapple_balanced_2025, beaulieu_fast_2026, wang_longitudinal_2025}. Numerical and analytical analysis shows that one can achieve leakage removal times approx.~\SI{30}{\nano\second} for the pumped and dissipated transmon system and \SI{125}{\nano\second} for the junction readout scheme. In both cases perturbations within the qubit subspace are minimized. 

\begin{figure}
  \centering
  \includegraphics[width=0.99\linewidth]{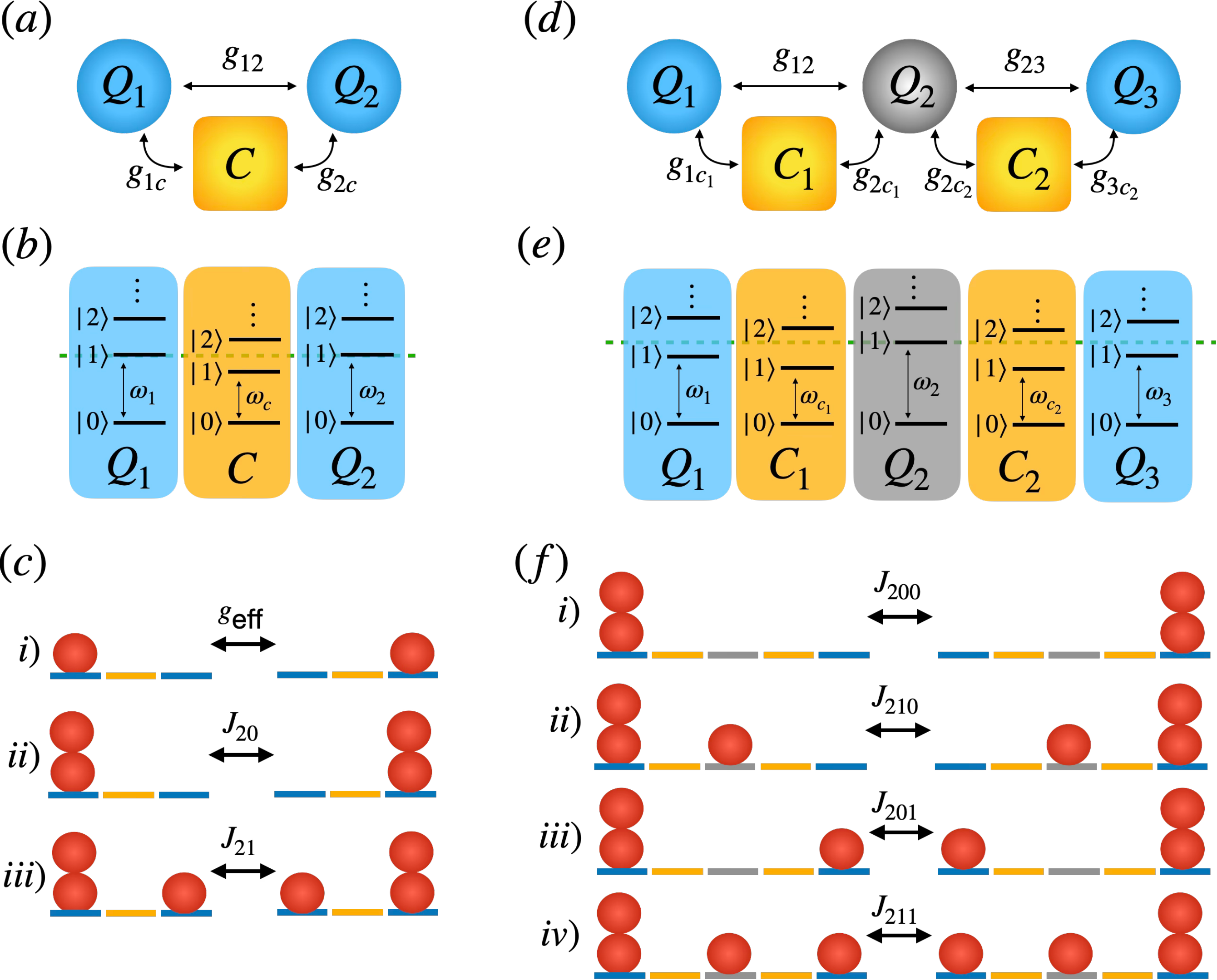}
  \caption{Schematics of the coupled transmon systems: Two transmons and one coupler~(a)-(c), and an array of coupled transmons~(d)-(f). (a), (d) Connectivity map showing the coupling configurations between the transmons $Q_j$ and the mediating tunable couplers $C_j$. (b),~(e)~Anharmonic multi-level frequency spectra for the transmons and couplers. The green dashed lines indicate the reference frequencies used to define the localized detunings, given by $\Delta_{jc} = \omega_j - \omega_c > 0$ for $j=1,2$ in panel~(b), and $\Delta_{c} = \omega_2 - \omega_{c_j} > 0$ for $j=1,2$ alongside $\Delta_{q} = \omega_2 - \omega_{j}$ for $j=1,3$ in panel~(e). (c), (f)~Conceptual illustration on the dynamics of leakage excitations (red circles) between the edge transmons (blue lines). Label~i) represents the qubit subspace oscillations governed by the effective hopping strength $g_{\text{eff}}$, while the remaining configurations depict distinct leakage mobility characterized by their respective effective hopping strengths.}
  \label{fig:scheme_dynamics}
\end{figure}

\begin{figure*}
  \centering
  \includegraphics[width=0.99\linewidth]{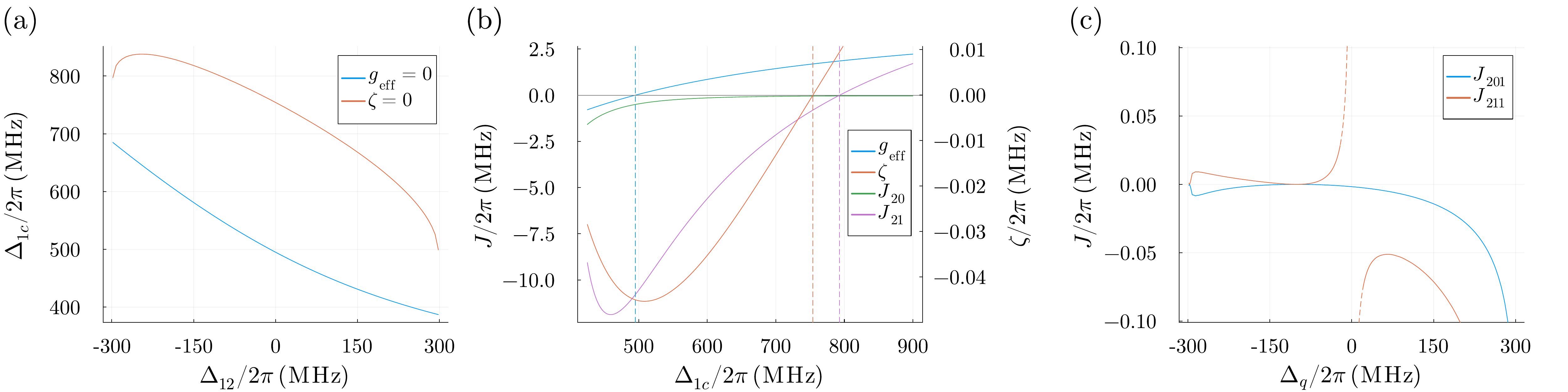}
  \caption{(a) Coupler detunings $\Delta_{1c} = \omega_1 - \omega_c$ where the excitation exchange $g_{\rm eff}$ or the $\textrm{ZZ}$ interaction is eliminated, $g_{\rm eff} = 0$ and $\zeta = 0$ as a function of the two transmon detuning $\Delta_{12}=\omega_1 - \omega_2$. (b) Effective coupling strengths for different coupler detuning~$\Delta_{1c}$: excitation exchange $g_{\mathrm{eff}}$ for $\ket{10}\!\leftrightarrow\!\ket{01}$, leakage exchange $J_{20}$ for $\ket{20}\!\leftrightarrow\!\ket{02}$ and $J_{21}$ for $\ket{21}\!\leftrightarrow\!\ket{12}$, and $\textrm{ZZ}$~coupling $\zeta$. (c) Effective leakage tunneling strengths $J_{201}$ for $\ket{201}\!\leftrightarrow\!\ket{102}$ and $J_{211}$ for $\ket{211}\!\leftrightarrow\!\ket{112}$, when the two edge transmons are detuned from the middle transmon by $\Delta_q$ and the couplers are tuned so that $\zeta = 0$. In panel (a) the second transmon is tuned to $\omega_2/2\pi = \SI{4000}{\mega\hertz}$, in panel (b) the two transmons are resonant $\omega_1/2\pi = \omega_2/2\pi = \SI{4000}{\mega\hertz}$, and in panel (c) the middle transmon is tuned to $\omega_2/2\pi = \SI{4000}{\mega\hertz}$. Here the transmon-transmon coupling is $g_{12} / 2\pi = \SI{5}{\mega\hertz}$ and the transmon-coupler coupling is $g_{1c} / 2\pi = -g_{2c} / 2\pi = \SI{50}{\mega\hertz}$. The transmons and the coupler have equal anharmonicity $U_1/2\pi = U_2/2\pi = U_c/2\pi = \SI{300}{\mega\hertz}$.}
  \label{fig:effective_coupling}
\end{figure*}

The paper is structured as follows. Section~\ref{sec:model} introduces the system of transmons and tunable couplers with employed analytical and numerical methods. Leakage mobility between two transmons and a tunable coupler is discussed in Sec.~\ref{sec:leakage.mobility}. Section~\ref{sec:leakage.tunneling} deepens the analysis to leakage tunneling between next-nearest transmons. Next, two leakage removal units based on leakage mobility are proposed in Secs.~\ref{Sec:lru}-\ref{sec:LRU2}. Summary and discussions are given in Sec.~\ref{sec:disc}.

\section{Transmons with tunable couplers}\label{sec:model}
\subsection{Two coupled transmons}
We first consider a system consisting of two computational transmons coupled directly with strength $g_{12}$ and via a central tunable transmon coupler (hereafter referred to as the coupler) connected to both transmons with strength $g_{jc}$, see Fig.~\ref{fig:scheme_dynamics}(a). The system dynamics is described by the Hamiltonian~\cite{yan_2018}
\begin{align}
    \frac{\hat{H}}{\hbar} = \sum_{\substack{j=1,2\\\ \ c}} \left[ \omega_j \hat{n}_j - \frac{U_j}{2}\hat{n}_j(\hat{n}_j - 1) \right] + \hat{V}_{12} + \sum_{j=1,2}\hat{V}_{jc},
    \label{H_full}
\end{align}
where $\hat{a}_j$ denotes the bosonic annihilation operator for the transmons $j=1,2$ and the coupler $j=c$, with $\hat{n}_j = \hat{a}_j^\dagger \hat{a}^{}_j$ representing their respective number operators. The parameters $U_j$ characterize the localized anharmonicity and $\omega_j$ is the independently tunable on-site frequency for each site. The interaction terms are defined as $\hat{V}_{ij} = g_{ij}( \hat{a}_i^\dagger \hat{a}^{}_j + \hat{a}^{}_i \hat{a}_j^\dagger)$ originating in capacitive coupling between the elements. When site dependence is not relevant, we refer to the transmon-transmon couplings as $g_q\equiv g_{ij}$ and to the transmon-coupler couplings as $g_c\equiv |g_{jc}|$. 

Our main focus is in characterizing the passive excitation dynamics between the transmons. This includes both the leakage state $\ket{2}$, as well as the qubit subspace states $\ket{0}$ and $\ket{1}$ [Fig.~\ref{fig:scheme_dynamics}(c)]. The system is operated in the dispersive, weakly coupled regime where the direct transmon-transmon interaction is much weaker than the transmon-coupler interaction, $|g_{12}| \ll |g_{jc}|$. Operating at the relevant points requires a large detuning between the transmons and the coupler. Given that the individual transmon-coupler interactions possess opposite signs, $g_{1c} = -g_{2c}$, this architecture requires a detuning where the coupler frequency is smaller than the transmon frequency, that is, a positive detuning $\Delta_{jc} = \omega_j - \omega_c > 0$~[Fig.~\ref{fig:scheme_dynamics}(a)]. While this configuration offers clear advantages regarding leakage effects~\cite{stehlik_2021} and scalability~\cite{eyob_2021}, it makes the analytic perturbative analysis more complicated. The early tunable coupler proposals~\cite{chen_2014, yan_2018, li_2020} used a negative transmon-coupler detuning.

 The Hamiltonian of Eq.~\eqref{H_full} is written in the rotating-wave approximation omitting the counter-rotating terms $\hat{a}_i\hat{a}_j$ and $\hat{a}_i^\dagger \hat{a}_j^\dagger$ in the coupling parts~\cite{magesan_2020, kandala_2021}. For the passive, non-driven dynamics considered here, these terms have a negligible effect. Their omission, however, ensures that the total excitation number is a conserved quantity reducing the dimensionality of the Hilbert space, simplifying the analytical derivations, and easing the numerical evaluations. We note that the counter-rotating terms must be retained when modeling active quantum gate operations or exploring highly driven regimes at specific operational biases~\cite{marxer_2023}.

 We will also explore the utility of this two transmons and a tunable coupler architecture as a passive leakage removal unit (LRU). Then the first transmon is designated as the primary computational qubit site, while we add engineered local dissipation to the second transmon to systematically remove the leaked population. For the analytical treatment of this open quantum system, we map the derived effective Hamiltonians onto the framework of Ref.~\cite{martinvazquez_2025}, which enables the evaluation of leakage-removal rates and population dynamics. Dissipative effects are characterized through the non-unitary evolution generated by the effective non-Hermitian Hamiltonian, with the decay of the state norm providing a measure of population loss from the effective subspace. In addition, non-Hermitian perturbation theory is employed to derive analytical estimates of the decay rates in different regimes~\cite{martinvazquez_2024}.

\subsection{Array of coupled transmons}
To investigate next-nearest-neighbor tunneling effects, we extend this architecture to a larger array comprising three transmons, where adjacent pairs are coupled both directly and via intermediate tunable couplers, schematized in Fig.~\ref{fig:scheme_dynamics}(d). The corresponding Hamiltonian is generalized as
\begin{align}
    \frac{\hat{H}}{\hbar} =& \sum_{\substack{j=1,2,3 \\ c_1,c_2}} \left[ \omega_j \hat{n}_j - \frac{U_j}{2}\hat{n}_j(\hat{n}_j - 1) \right] + \hat{V}_{12}  +\hat{V}_{23}\notag \\
     &+\sum_{j=1,2}\hat{V}_{jc_1} + \sum_{j=2,3}\hat{V}_{jc_2},
    \label{H_tunneling_full}
\end{align}
where $j=1,2,3$ are the transmons and $j=c_1,c_2$ the tunable couplers. In this case, we will describe the dynamics of the leakage excitations between the two edge transmons $j=1,3$, see Fig.~\ref{fig:scheme_dynamics}(f). In this architecture, we define the detuning of the coupler with respect the middle transmon $ \Delta_c=\omega_{2}-\omega_{c_j}>0$ for $j=1,2$ [Fig.~\ref{fig:scheme_dynamics}(e)]. We also define the detuning between the middle transmon and the edge transmons as $\Delta_q=\omega_{2}-\omega_{j}$ for $j=1,3$.

The dynamics of the coupled transmon systems are studied here through a combination of analytical and numerical methods. We use the QuantumToolbox.jl~\cite{QuantumToolbox.jl2025} package to numerically solve the time evolution of the system. In Sec.~\ref{Sec:lru} and Sec.~\ref{sec:LRU2}, the package is used to simulate the dissipative dynamics of the leakage removal schemes, including the driving of the pumped scheme. To gain physical insight into the leakage dynamics mechanisms, we derive an effective Hamiltonian for the transmons by perturbatively decoupling the couplers via a Schrieffer-Wolff transformation~\cite{bravyi_2011}. Operating within the dispersive regime, the small perturbative parameter is  $\lambda \equiv |g_{jc}/\Delta_{jc}| \ll 1$. Because our analysis extends beyond the computational subspace to capture higher-order leakage processes, we must explicitly account for the other scaling terms that depend on anharmonicity, such as $g_{jc}/|\Delta_{jc} \pm U|$ and $g_{ij}/U$. While these quantities are small and assumed to scale on the order of $\lambda$, tracking their variations across the parameter space is essential to guarantee the quantitative validity and reliability of the perturbative framework. While our focus is in the leakage excitation dynamics, we note that the leakage subspace effects have been studied earlier but in the context of their effects on $\textrm{CZ}$-gate implementation, see, \textit{e.g.}, Ref.~\cite{espinos_2023}.

\section{Leakage dynamics in tunable-coupled transmons} \label{sec:leakage.mobility}
We first examine a system comprising two transmons connected via a tunable coupler, Eq.~\eqref{H_full}. The goal is to characterize the differential dynamics within the single-excitation subspace and evaluate leakage dynamics effects, see Fig.~\ref{fig:scheme_dynamics}(a). Truncating the Schrieffer-Wolf expansion at the second order in $\lambda$, we obtain an explicit effective Hamiltonian for the two-transmon system. The coupler has been traced out from the dynamics. Its state is characterized only via the fixed coupler excitation number~$n_c$. The state $\ket{n k}$ refers now only to the transmons. The effective Hamiltonian is given by:
\begin{align}
\hat H^{(2)}
=&
\sum_{j=1,2}
\hbar \left[\omega_j \hat n_j - \frac{U_j}{2}\hat n_j(\hat n_j - 1) \right] \nonumber \\
&+\hbar g_{12}\left( \hat a_1^\dagger \hat a^{}_2 + \hat a_2^\dagger \hat a_1^{} \right) + \hat{E}^{(2)}+\hbar \hat{J}^{(2)},
\label{H_eff}
\end{align}
where the energy $\hat E^{(2)} $ and hopping $\hat J^{(2)}$ corrections are given by 
\begin{align}
\hat{E}^{(2)}&=\sum_{j=1,2} \hbar \left [
\frac{(n_c+1)g_{jc}^2\hat{n}_j}{\Delta_{jc}-(\hat{n}_j-1)U_j+n_cU_c} \right. \nonumber \\
& \left. \qquad \qquad \quad -\frac{n_cg_{jc}^2(\hat{n}_j+1)}{\Delta_{jc}-\hat{n}_jU_j+(n_c-1)U_c}\right ],\\
\hat{J}^{(2)}&=\frac{g_{1c}g_{2c}}{2}\hat{a}_1^\dagger \hat{a}^{}_2
\left[  \frac{(n_c+1)}{\Delta_{1c}-\hat{n}_1U_1+n_cU_c} \right. \nonumber \\
&\qquad \qquad \quad  +\frac{(n_c+1)}{\Delta_{2c}-(\hat{n}_2-1)U_2+n_cU_c}  \label{eq.J2eff} \\
& \qquad \qquad \quad - \frac{n_c}{\Delta_{2c}-(\hat{n}_2-1)U_2+(n_c-1)U_c} \nonumber\\ 
&\left.\qquad \qquad \quad -\frac{n_c}{\Delta_{1c}-\hat{n}_1U_1+(n_c-1)U_c}
\right] + \text{h.c.}. \nonumber
\end{align}
Notice that the correction terms $\hat E^{(2)} $ and $\hat J^{(2)}$ contain dependence on transmon population $\hat n_j$ both in nominators and denominators rendering different dynamics for example between the states $\{\ket{10},\ket{01}\}$ than  $\{\ket{20},\ket{02}\}$. Higher order corrections for Eq.~\eqref{H_eff} are included in App.~\ref{App:couplers}. 

Tunable couplers provide a versatile architecture for controlling qubits interactions, enabling the selective enhancement or suppression of specific system dynamics. Two operational regimes are of particular importance: the cancellation of the effective single-excitation exchange coupling $g_{\text{eff}} = 0$ and the elimination of the residual $\textrm{ZZ}$ interaction $\zeta = 0$ [Fig. \ref{fig:effective_coupling}(a)]. The $g_{\text{eff}} = 0$ regime corresponds to the suppression of $\ket{10} \leftrightarrow \ket{01}$ coherent exchange dynamics [case $i)$ in Fig.~\ref{fig:scheme_dynamics}(c)]. This condition is achieved via the destructive interference of multiple coupling paths, described to the leading order by $g_{\text{eff}}=g_{12}+g_{1c}g_{2c}( \Delta_{1c}^{-1}+\Delta_{2c}^{-1})/2$~\citep{yan_2018} with higher-order corrections derived up to the fourth order in App.~\ref{App:couplers}. Conversely, the $\textrm{ZZ}$-OFF condition $\zeta = 0$ denotes the suppression of parasitic longitudinal interactions between the transmons, effectively canceling the dispersive shift term $\hat{n}_1 \hat{n}_2$ appearing in the higher order version of Eq.~\eqref{H_eff}. The precise frequency for the cancellation of the $\textrm{ZZ}$ interaction is typically determined numerically from the eigenenergies of Eq.~\eqref{H_full} as $\zeta = (E_{101} - E_{100} - E_{001} + E_{000})/\hbar $~\cite{stehlik_2021, marxer_2023}, where $E_{ijk}$ are eigenenergies of the transmons and couplers. The condition can also be analytically approximated using the fourth-order perturbation theory~\cite{chu_2021, sung_2021}. Achieving the $\zeta = 0$ operational point requires the coupler to be more further detuned from the transmons frequencies than it is for the single-excitation exchange cancellation condition.

While the condition $g_{\text{eff}} = 0$ eliminates exchange dynamics within the computational subspace, the effective coupling between the computational and leakage manifolds, as well as between states within the leakage subspace remains non-zero [Fig.~\ref{fig:effective_coupling}(b)]. Technically this originates from the effective hopping strength $\hat J^{(2)}$ term being dependent on the transmon occupation operators~$\hat n_i$. Physically, the coupling paths for higher excitations get different phases and weightings not resulting same destructive interference conditions as withing the single excitation manifolds. This is one of the core results of our work. 

To characterize leakage dynamics effects, we examine the effective Hamiltonian in the presence of leakage states such as $\ket{20}$ and $\ket{21}$ [cases $ii)$ and $iii)$ in Fig.~\ref{fig:scheme_dynamics}(c), respectively]. Due to the particle conservation, we can model the situations as effective few-level systems governed by interaction strengths $J_{\text{eff}}$ and energy detunings $\Delta E$, which together dictate the population transfer between transmons in each case. Our analysis focuses on the leading-order contributions of each term. In App.~\ref{App:couplers}, we provide the full analytical and algebraically more complex expressions for the higher-order terms.

\begin{figure*}
  \centering
  \includegraphics[width=\linewidth]{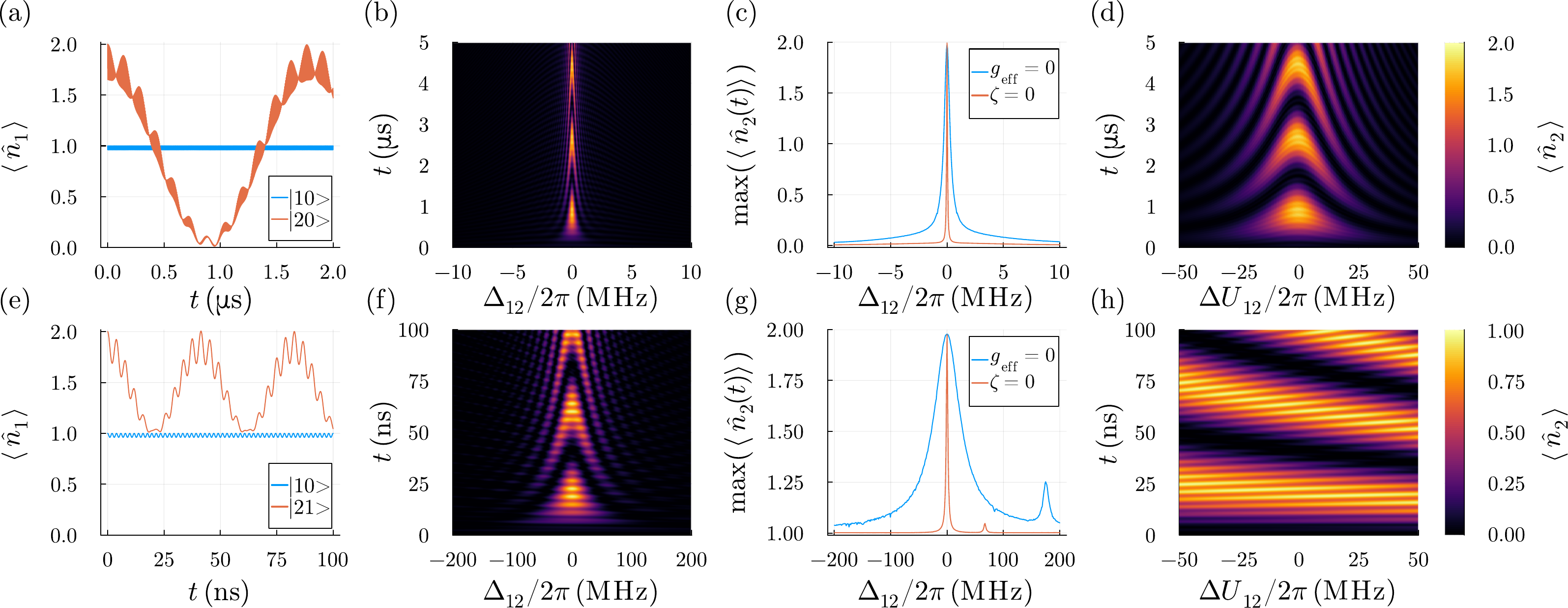}
  \caption{(a), (e) Leakage time dynamics for two resonant transmons when the coupler is tuned so that $g_{\rm eff} = 0$. (b), (f) Sensitivity of the leakage transfer to the frequency detuning between the first and second transmons. (c), (g) Maximum leakage transfer between the two transmons while the coupler is tuned so that $g_{\rm eff} = 0$ or $\zeta = 0$. (d), (h) Sensitivity of the leakage transfer to differences in anharmonicity between the transmons while the relevant transition is kept on-resonance. In the top row the initial state is $\ket{\psi_0} = \ket{20}$ and in the bottom row $\ket{\psi_0} = \ket{21}$. We assume the transmons have identical anharmonicities, so that the detuning of the second transmon in (b) and (f) directly translates to a detuning between the relevant transitions. In (c) and (g) the maximum population transfer is calculated from \SI{10}{\micro \second} and \SI{3}{\micro \second} of time-evolution, respectively. In the simulations we use the same parameters as in Fig.~\ref{fig:effective_coupling}, with $\omega_c/2\pi = \SI{3505}{\mega\hertz}$. In (b)-(d) and (f)-(h) where the transmons are off-resonance, we use the analytical solution from Ref.~\onlinecite{yan_2018} with a constant correction to determine the coupler tuning $\omega_c$ so that $g_{\rm eff} = 0$.}
  \label{fig:leakage_dynamic}
\end{figure*}

\subsection{Leakage in the first transmon with the second transmon in the ground state}\label{Sec:coupler_200}
Figure~\ref{fig:leakage_dynamic}(a)-(d) shows the dynamics in the case where a leakage excitation is present in the first transmon while the second remains in the ground state, resulting in effective $\ket{20} \leftrightarrow \ket{02}$ dynamics [case $ii)$ in Fig.~\ref{fig:scheme_dynamics}(c)]. Here we select an operation point where qubit subspace dynamics between the transmons is suppressed $g_{\text{eff}} = 0$. The leakage excitation transition from a transmon to an other occurs at fourth order, mediated by three distinct pathways: fourfold hopping through the coupler~$\sim \lambda^3 g_c$, twofold direct hopping between transmons~$\sim g_{12}^2/U$, and a combined mechanism~$\sim \lambda^2 g_{12}$. Consequently, these dynamics are inherently slow, the leakage hops to the neighboring site in approx.~\SI{1}{\micro\second}, see Fig.~\ref{fig:leakage_dynamic}(a). The slow dynamics is sensitive to the transmon detuning $\Delta_{12}$, as the energy mismatch $\Delta E \sim 2 \Delta_{12}$ can easily dominate the small effective coupling $J_{20} \sim \lambda^3 g_c$~[Fig.~\ref{fig:leakage_dynamic}(b)]. When relaxing the leakage level resonance condition [Fig.~\ref{fig:leakage_dynamic}(b)], the width of the chevron pattern reveals the coupling strength which is of the order of \SI{0.20}{\mega\hertz}. While previous case assumed the suppression of qubit subspace dynamics $g_{\text{eff}} = 0$, Fig.~\ref{fig:leakage_dynamic}(c) shows the frequency detuning dependence under the condition $\zeta = 0$. This reveals a enhanced sensitivity to frequency mismatch, as the $\zeta = 0$ regime coincides with the parameters that reduce the magnitude of $J_{20}$, see Fig.~\ref{fig:effective_coupling}(b). 

An interesting question is whether it is possible to find a condition when the $\ket{20} \leftrightarrow \ket{02}$ transition would be fully suppressed. It turns out that this dynamics cannot be exactly blocked, see details in App.~\ref{App:couplers}. Due to the high-order nature of the system dynamics, lower-order processes must be carefully accounted for, specifically the leakage splitting transitions $\ket{20} \leftrightarrow \ket{11}$. The splitting dynamics become resonant when $\omega_2 = \omega_1 - U_1$, with an effective hopping strength given by  $J_{20,11}=\sqrt{2}\left\{g_{12}+g_{1c}g_{c2}\left[\Delta_{1c}^{-1}+(\Delta_{1c}-U_1)^{-1}\right]/2\right\}$. Note that this interaction is asymmetric with respect to leakage in the second transmon; for the transition $\ket{02} \leftrightarrow \ket{11}$, the resonance condition shifts to $\omega_2 = \omega_1 + U_2$, and the coupling strength becomes $     J_{02,11}=\sqrt{2}\left\{g_{12}+g_{1c}g_{c2}\left[\Delta_{1c}^{-1}+(\Delta_{2c}-U_2)^{-1}\right]/2\right\}$. These transitions are equivalent to the ones described in Refs.~\cite{espinos_2023, sanclemente_2026}; however, setting them to zero does not ensure the blocking of higher order dynamics through the coupler as described in \eqref{J_20_4th}. Nevertheless, by operating far from these resonances, a large transmon-coupler detuning effectively ensures that leakage transitions are practically suppressed [Fig. \ref{fig:effective_coupling}(b)]. 

We now study the sensitivity of the transition to parameter variations. First, by imposing the resonance condition for leakage excitations $2\omega_1 - U_1 = 2\omega_2 - U_2$, we examine the dependence on the anharmonicity detuning $\Delta U_{12}=U_1-U_2$ between the transmons. We observe that while a dependence exists, the system is quite weakly sensitive to variations in this parameter, see Fig.~\ref{fig:leakage_dynamic}(d). Interestingly, even under this resonance condition, inhomogeneous second-order energy corrections arise when the bare frequencies are off-resonance $\omega_1 \neq \omega_2$. This results in an energy mismatch $\Delta E \sim \lambda^2 \Delta U_{12}$ that remains larger than the fourth-order hopping $J_{20}$, explaining the observed sensitivity.

\subsection{Leakage in the first transmon with the second transmon in the single-excitation state}\label{Sec:coupler_201}
Figure~\ref{fig:leakage_dynamic}(e)-(h) depicts the dynamics for the setup of a leakage excitation in the first transmon and a single excitation in the second transmon. Again here we assume that the condition $g_{\rm eff}=0$. The effective $\ket{21} \leftrightarrow \ket{12}$ dynamics [case $iii)$ in Fig.~\ref{fig:scheme_dynamics}(c)] occurs now at second order involving two coupler-mediated hoppings~$\sim \lambda g_{jc}$ and a single direct transmon-transmon hopping~$g_{12}$. This leads to much faster dynamics, where the leakage propagates between the transmons in~\SI{20}{\nano\second}, see Fig.~\ref{fig:leakage_dynamic}(e). Due to the stronger effective coupling [Fig.~\ref{fig:effective_coupling}(b)], the system is less sensitive to the transmon detuning $\Delta_{12}$, as the hopping $J_{21} \sim \lambda g_c$ is more robust against variations in the energy mismatch $\Delta E \sim \Delta_{12}$ [Fig.~\ref{fig:leakage_dynamic}(f)]. As with the previous case, Fig.~\ref{fig:leakage_dynamic}(g) shows that the $\zeta = 0$ condition results in a stricter frequency detuning dependence than the $g_{\text{eff}} = 0$ condition.

Furthermore, we find a condition for blocking the dynamics for the $\ket{21} \leftrightarrow \ket{12}$ transition, given by
\begin{align}
    J_{21}=2g_{12}+g_{1c}g_{c2}\left(\frac{1}{\Delta_{2c}-U_2}+\frac{1}{\Delta_{1c}-U_1}\right)=0.
    \label{201_blocking_2nd_order}
\end{align}
This is actually located near the $\zeta = 0$ condition as seen in Fig.~\ref{fig:effective_coupling}(b), and on the narrow peak in Fig.~\ref{fig:leakage_dynamic}(g). Fourth order corrections to the leakage dynamics are given in App.~\ref{App:couplers}.

When the leakage resonance condition is maintained $\omega_1 - U_1 = \omega_2 - U_2$, the dynamics become largely independent of the anharmonicity detuning [Fig.~\ref{fig:leakage_dynamic}(h)], because the energy difference $\Delta E \sim 2\lambda^2 \Delta U_{12}$ is small compared to $J_{21}\sim \lambda^2 (U_1+U_2)$. Notably, an increase in the oscillation frequency is observed for positive anharmonicity detuning, scaling as $\Omega \sim 2\lambda^2 \sqrt{\Delta U_{12}^2+(U_1+U_ 2) ^2}$. This behavior can be explained by the increase of the effective hopping rate: as one of the energy denominator $\Delta_{jc} - U_j$ decreases, the corresponding virtual transitions are faster. A qualitatively similar but more subtle effect occurs at higher order, as seen in Fig.~\ref{fig:leakage_dynamic}(d). 

\begin{figure*}
  \centering
  \includegraphics[width=\linewidth]{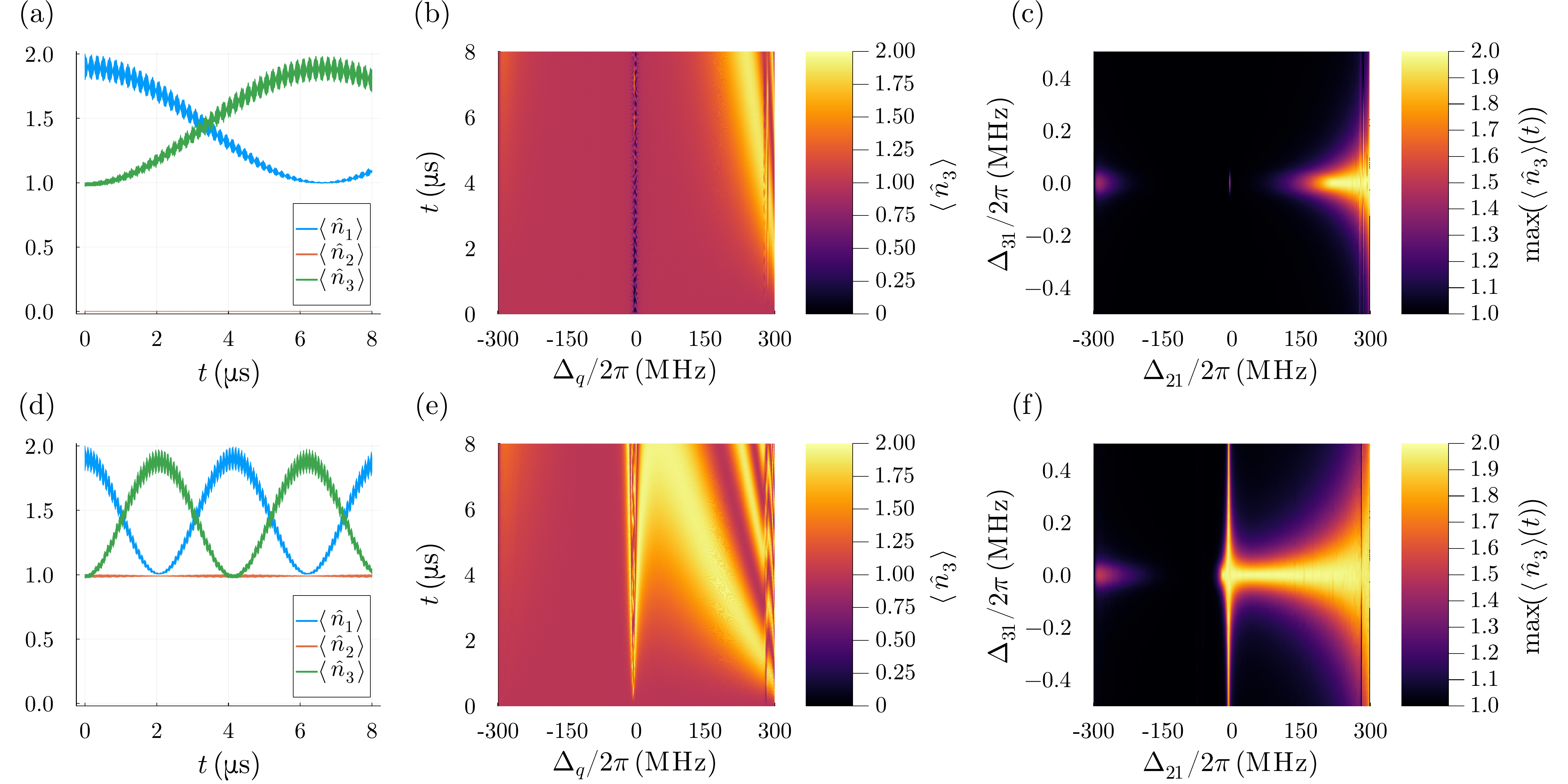}
  \caption{Leakage tunneling in a chain of three transmons, where the edge transmons are resonant and the middle transmon is detuned by $\Delta_q = \omega_2 - \omega_1 = \omega_2 - \omega_3$. The couplers are tuned so that $\zeta = 0$. In (a)-(c) the initial state is $\ket{\psi_0} = \ket{201}$. In (d)-(f) the initial state is $\ket{\psi_0} = \ket{211}$. (a), (d) Time evolution of leakage excitation. (b), (e) Time evolution of leakage excitation for different detunings $\Delta_q$. (c), (f) Sensitivity of the tunneling to the off-resonance of the edge transmons $\Delta_{31} = \omega_3 - \omega_1$. For the anharmonicities, transmon-transmon coupling and transmon-coupler coupling we use same parameters as in Fig.~\ref{fig:effective_coupling}. The middle transmon is kept at $\omega_2/2\pi = \SI{4000}{\mega\hertz}$ and the couplers tuned so that $\zeta = 0$ between the first and second, and second and third transmon. In (a) and (d) we use $\omega_1/2\pi = \omega_3/2\pi = \SI{3750}{\mega\hertz}$ and $\omega_{c1}/2\pi = \omega_{c2}/2\pi = \SI{3162.762}{\mega\hertz}$. In (c) and (f) the maximum population transfer is calculated from \SI{10}{\micro \second} of time-evolution.}
  \label{fig:tunneling_dynamics}
\end{figure*}

\section{Leakage Tunneling in tunable-coupled transmon arrays}\label{sec:leakage.tunneling}
Here we are interested in the scenario where a larger transmon-coupler array is set in a $\textrm{ZZ}$-OFF condition, modeling an idling phase of quantum processor. Thus, we consider the system of three transmons coupled via two couplers [Fig.~\ref{fig:scheme_dynamics}(d)] tuned so that $\zeta = 0$. Although as demonstrated in the preceding section that leakage propagation is suppressed in this frequency tuning, the dynamics become more complex when considering an array of transmons. Specifically, we consider a chain of three transmons where the central transmon is off-resonance relative to the others, while the edge transmons remain resonant forming an ABA configuration. In this section, we investigate the conditions under which leakage propagates via tunneling between the edge transmons.

In App.~\ref{App:array_couplers}, we detail the derivation of the effective Hamiltonian up to the second order including relevant higher-order corrections following the perturbative approach established in Sec.~\ref{sec:leakage.mobility}. A key distinction in this configuration is that the central transmon receives a unique second-order energy correction. This induces an effective detuning relative to the edge transmons, even in the case of homogeneous physical parameters. 

Despite the system dynamics across various parameter configurations and excitation manifolds is complex, we identify the configurations most relevant to leakage mobility via perturbative analysis. For instance, we neglect the tunneling processes $\ket{200} \leftrightarrow \ket{002}$ and $\ket{210} \leftrightarrow \ket{012}$ [cases $i)$ and $ii)$ in Fig.~\ref{fig:scheme_dynamics}(f), respectively], as $J_{200}$ and $J_{210}$ scale both as $\sim  \lambda^7 g_c$ and are thus considered negligible weak for practical purposes. Instead, this section focuses on the tunneling dynamics between $\ket{201} \leftrightarrow \ket{102}$ and $\ket{211} \leftrightarrow \ket{112}$ [cases $iii)$ and $iv)$ in Fig.~\ref{fig:scheme_dynamics}(f), respectively]. The effective hopping strengths of theses processes $J_{201}$ and $J_{211}$ occur at order $\sim \lambda^3 g_c$, with variations in rate arising from differing virtual hopping pathways [Figs.~\ref{fig:tunneling_dynamics}(a),(d)]. 

The detuning between the central transmon and the couplers is defined as $\Delta_{c} = \omega_2 - \omega_c$ and the detuning between the central and the edge transmons is $\Delta_q = \omega_2 - \omega_1 = \omega_2 - \omega_3$ assuming that the edge transmons are on resonance for the initial characterization. Furthermore, we assume a uniform transmon-transmon coupling strength $g_q = g_{12} = g_{23}$, staggered coupler-transmon coupling strengths $g_c = g_{1c_1}=-g_{2c_1} =-g_{2c_2}= g_{3c_2}$, and homogeneous anharmonicity $U$ across the system. While the analytical framework presented below maintains a general parameterization, our primary focus is the $\textrm{ZZ}$-OFF operating condition, which requires a unique coupler detuning $\Delta_c$ for each specific transmon detuning $\Delta_q$, as it is shown in Fig. \ref{fig:effective_coupling}(a). 

\subsection{Leakage in the first transmon with the third transmon in the single-excitation state}
The dynamics of the $|201\rangle \leftrightarrow |102\rangle$ transition [case $iii)$ in Fig.~\ref{fig:scheme_dynamics}(f)] are characterized by two primary resonance conditions defined relative to the specific system dynamics [Fig.~\ref{fig:tunneling_dynamics}(b)]. First, resonance occurs at $\Delta_q = \Delta_{c} - U$, where the edge transmons are resonant with the couplers for the $\ket{20_c}\leftrightarrow \ket{11_c} $ transition. In this regime, the dynamics are predominantly mediated through the couplers, with the effective hopping rate scaling as $J_{201} \sim g_c^4 / [\Delta_{c} (\Delta_{c} - \Delta_q - U)^2]$. Consequently, the interaction strength increases as the system approaches this point. Second, a resonance point exists at $\Delta_q = -U$, where all three transmons are on resonance for the transitions $\ket{\sfrac{20}{02}} \leftrightarrow \ket{11}$. Here, the dynamics are mediated via the middle transmon, and $J_{201} \sim g_q^2 / [\Delta_q + U +\mathcal{O}(\lambda g_c)]$. This resonant point marks the boundary of the $\textrm{ZZ}$-OFF condition (limited by $U/2\pi = 300~\text{MHz}$), where the effective hopping strength vanishes $J_{20,11} \approx 0$, resulting in a sharp suppression near the edge of the detuning [more noticeable in Fig.~\ref{fig:tunneling_dynamics}(c)]. Notably, when the transmons are tuned to exact resonance $\Delta_q = 0$, the system exhibits single-excitation oscillations  $\ket{201} \leftrightarrow \ket{210}$ that remains unblocked when operating in the $\textrm{ZZ}$-OFF condition.

As demonstrated, the relationship between leakage dynamics and the underlying spectral structure is non-trivial. In the regime considered here, where the coupler frequency is lower than that of the transmons $\Delta_c > 0$, different propagation scenarios emerge. Within the ABA configuration, this transition exhibits coupler-mediated dynamics when the central transmon frequency is higher than the edge transmon frequencies $\Delta_q>0$. Conversely, transport is primarily mediated by the central transmon when its frequency is lower than that of the edge transmons $\Delta_q<0$. In all scenarios, the resonance points are shifted by second-order energy corrections on the order of $\mathcal{O}(\lambda g_c)$. The sensitivity of these transitions to edge-transmon detuning can be understood through the analysis presented in Sec.~\ref{sec:leakage.mobility}. Analogous to the $\ket{21} \leftrightarrow \ket{12}$ case depicted in Fig.~\ref{fig:leakage_dynamic}(g), the effective two-state dynamics for this transition is confined within the same anharmonicity manifold. Consequently, a smooth reduction in the tunneling amplitude is observed as the detuning increase [Fig.~\ref{fig:tunneling_dynamics}(c)].

Despite the complexity of this parameter space, the conditions required to suppress leakage tunneling in this particular configuration can be derived by analyzing the effective hopping terms, see App.~\ref{App:array_couplers}. For this specific transition, the suppression condition is approximately $\Delta_q \approx \Delta_c  - U - g_c^2/g_q$ [Fig.~\ref{fig:map_tunneling_zeros}(a)]. We find that the regime for minimized mobility is relatively broad; for a representative anharmonicity of $U/2\pi = \SI{300}{\mega\hertz}$, this suppression condition is satisfied at $\Delta_q/2\pi \approx -\SI{101}{\mega\hertz}$ [Fig. \ref{fig:effective_coupling}(c)], showing excellent agreement with our numerical results [Fig.~\ref{fig:tunneling_dynamics}(b)]. 

\subsection{Leakage in the first transmon with the second and third transmon in the single-excitation state}
In contrast, the $|211\rangle \leftrightarrow |112\rangle$ transition [case $iv)$ in Fig.~\ref{fig:scheme_dynamics}(f)] exhibits distinct behavior due to the occupancy of the central transmon [Fig.~\ref{fig:tunneling_dynamics}(e)]. In the coupler-mediated regime, the resonance condition is $\Delta_q = \Delta_{c} - U$, and the effective hopping include also terms that scale as $J_{211} \sim g_c^4 / [(\Delta_{c} - U) (\Delta_{c} - \Delta_q - U)^2]$. The presence of the $(\Delta_{c} - U)$ term indicates that these dynamics are faster than the $|201\rangle \leftrightarrow |102\rangle$ case, as the virtual hopping processes occur within the same anharmonicity manifold. For the middle-transmon-mediated case, there are now two resonance conditions involving the dynamics $\ket{211}\to\ket{121}\to\ket{112} $ and $\ket{211}\to\ket{202}\to\ket{112} $. For the first case, $J_{211}\sim g_q^2 / [\Delta_q + \mathcal{O}(\lambda g_c)]$, while for the second case, $J_{211} \sim g_q^2 / [\Delta_q +U+ \mathcal{O}(\lambda g_c)]$ and it shows the same behavior as $\ket{201} \leftrightarrow \ket{102} $ near the edge [see Fig.~\ref{fig:tunneling_dynamics}(f)]. 

For the $\ket{211} \leftrightarrow \ket{112}$ leakage channel, several distinct mobility scenarios emerge when the coupler frequency is lower than that of the transmons $\Delta_c > 0$. Similarly, in the standard ABA configuration, the system exhibits coupler-mediated dynamics when the central transmon frequency is higher than the edge transmon frequencies $\Delta_q>0$, while transport is primarily mediated by the central transmon when $\Delta_q<0$. However, an additional central-transmon-mediated channel appears near the near-resonant BBB configuration. As with other channels, the exact resonance points across these scenarios are shifted by second-order energy corrections on the order of $\mathcal{O}(\lambda g_c)$. While fundamentally analogous to the $\ket{21} \leftrightarrow \ket{12}$ case, see Fig.~\ref{fig:leakage_dynamic}(g), this transition exhibits a significantly lower sensitivity to edge-transmon detuning [Fig.~\ref{fig:tunneling_dynamics}(f)] compared to the $\ket{201} \leftrightarrow \ket{102}$ case [Fig.~\ref{fig:tunneling_dynamics}(c)]. This enhanced resilience is related to the nature of the underlying virtual processes. Whereas the $\ket{211} \leftrightarrow \ket{112}$ virtual transitions are largely confined within a single anharmonicity manifold, thereby decoupling it from inter-manifold variations, the $\ket{201} \leftrightarrow \ket{102}$ virtual transitions rely on virtual hopping pathways between different anharmonicity manifolds, which amplifies its sensitivity to detuning.

Due to the complicated structural dependencies and multiple resonance conditions included in the effective hopping strength for the $\ket{211} \leftrightarrow \ket{112}$ tunneling dynamics, an analytical  simple approximation for the blocking condition, as the one derived for the $\ket{201} \leftrightarrow \ket{102}$ transition, cannot be straightforwardly obtained [Fig.~\ref{fig:map_tunneling_zeros}(b)]. Nevertheless, away from the localized resonance at $\Delta_q = 0$, both dynamics are qualitatively analogous. Specifically, the fourth-order coupler-mediated interactions differ only by the $(\Delta_c - U)^{-1}$ and $(\Delta_c - U)^{-2}$ dependencies that appears in the $\ket{211} \leftrightarrow \ket{112}$ transition, as seen by comparing Eqs.~\eqref{J_4_20001} and~\eqref{J_4_20101}. This similarity allows us to employ the previously established approximation $\Delta_q \approx \Delta_c - U - g_c^2/g_q$. This solution is illustrated in Fig.~\ref{fig:map_tunneling_zeros}(b), revealing a more strict blocking requirement than in the in the previous case. This approximation is less accurate as $\Delta_q$ approaches zero, where the condition derived in Sec.~\ref{Sec:coupler_201} becomes dominant, and for $\Delta_q < 0$ as the system approach the internal resonance point $\Delta_c = U$ of the coupler-mediated effective hopping, as mentioned above. Then, we get the same prediction of $\Delta_q/2\pi \approx -101~\text{MHz}$ [Fig. \ref{fig:effective_coupling}(c)], which is in excellent agreement with the numerical data presented in Fig.~\ref{fig:tunneling_dynamics}(e).

As a final remark on the tunneling blocking condition derived above, this section analyzed a specific ABA architecture, demonstrating that for leakage localized in transmon~A, setting $\omega_{\text{A}}/2\pi = 4101$~MHz and $\omega_{\text{B}}/2\pi = 4000$~MHz effectively suppresses tunneling to the adjacent $A$ transmons. However, in a realistic, extended array of the form ABAB, leakage originating within a type-B transmon would not be blocked. This limitation arises because the zero  condition is asymmetric with respect to $\Delta_q = 0$, leading to tunneling to adjacent B sites [Fig.~\ref{fig:scheme_dynamics}(i) and solid lines in Fig.~\ref{fig:map_tunneling_zeros}(c)]. Nevertheless, significant opportunities remain to mitigate tunneling within this generalized scenario. Beyond the straightforward approach for $1$D cases of introducing frequency detuning between the edge transmons as illustrated in Figs.~\ref{fig:tunneling_dynamics}(c,f), such as implementing an ABCABC... configuration, alternative mitigation strategies can be successfully devised. Upon examining the zero tunneling condition, it is evident that there are other relevant terms: the transmon anharmonicity $U$ and the ratio of the bare coupling strengths $g_c^2/g_q$. We show that either reducing the anharmonicity $U$ [black dashed line in Figs.~\ref{fig:map_tunneling_zeros}(a,b)] or increasing the ratio $g_c^2/g_q$ [solid black line in Figs.~\ref{fig:map_tunneling_zeros}(e,f)] yields a significantly more symmetric coupling profile with heavily suppressed effective hopping strengths. For instance, at a detuning of $40$~MHz, the tunneling oscillations across the entire array are mitigated to $\sim 5$~kHz by lowering the anharmonicity to $U/2\pi = 135$~MHz [dashed colored lines in Fig.~\ref{fig:map_tunneling_zeros}(c)]. Alternatively, configuring the system parameters to $g_c/2\pi \approx 63.25$~MHz and $g_q/2\pi = 4$~MHz provides even stronger protection, reducing the tunneling to $0.7$~kHz [Fig.~\ref{fig:map_tunneling_zeros}(g)].

\section{Tunable-Coupler-Enabled Passive Leakage Removal}\label{Sec:lru}
\begin{figure*}
  \centering
  \includegraphics[width=0.85\linewidth]{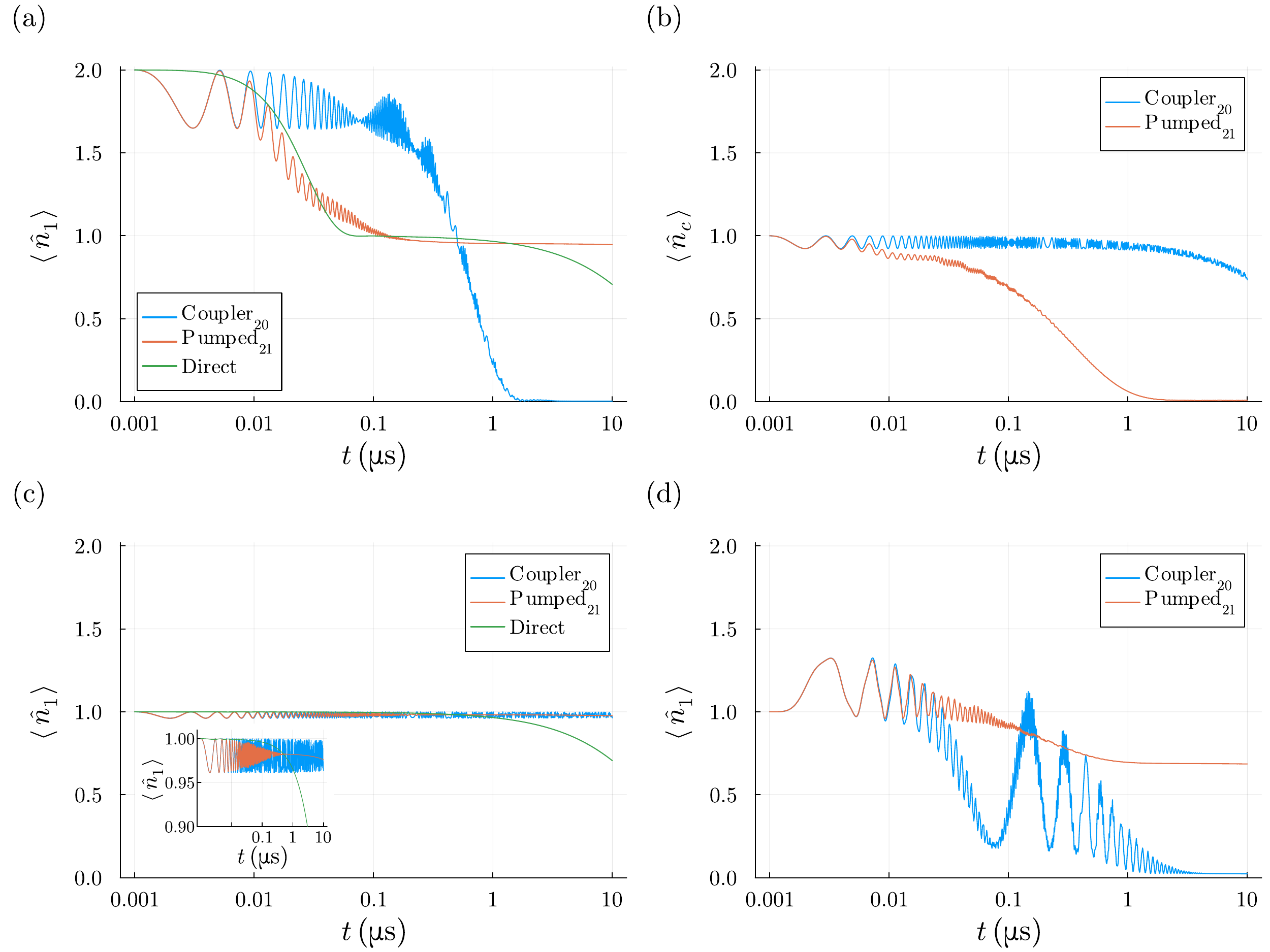}
  \caption{Leakage removal performance of the three leakage removal schemes: $\text{Coupler}_{20}$, $\text{Pumped}_{21}$ and Direct. (a) Leakage dynamics for the initial state $\ket{20}$. (b) Leakage dynamics with leakage initially in the coupler, $\ket{01_c0}$. (c) Qubit dynamics for the initial state $\ket{10}$. (d) Qubit dynamics with leakage initially in the coupler, $\ket{11_c0}$. Leakage in the coupler affects the effective coupling so that $g_{\rm eff} \neq 0$. For the Direct scheme we use the following parameters: $\omega_1/2\pi = \SI{4000}{\mega\hertz}$, $\omega_2/2\pi = \SI{3700}{\mega\hertz}$, $U/2\pi = \SI{300}{\mega\hertz}$, $J/2\pi = \SI{5}{\mega\hertz}$ and $\Gamma/2\pi = \SI{20}{\mega\hertz}$. For the tunable-coupler schemes we use the same parameters as in Fig.~\ref{fig:effective_coupling}, and the two transmons are resonant $\omega_1 = \omega_2$. In the $\text{Coupler}_{20}$ scheme the dissipation is $\Gamma/2\pi = \SI{0.4}{\mega\hertz}$. In the $\text{Pumped}_{21}$ scheme the dissipation is $\Gamma/2\pi = \SI{9}{\mega\hertz}$. \label{fig:lru_comparison}}
\end{figure*}
As discussed in Sec.~\ref{sec:leakage.mobility}, a tunable coupler can be used to cancel single-excitation exchange while simultaneously allowing leakage to propagate between the transmons. This suggests that the tunable coupler can be used to realize a passive leakage removal unit. We study tunable-coupler-enabled passive leakage removal by considering the first transmon as the coding qubit and the second transmon together with the tunable coupler as the leakage removal unit. For removing leakage from the system, we introduce dissipation to the second transmon. The two transmons are tuned such that the all energy levels are on resonance, and the tunable coupler is tuned to cancel the single-excitation exchange $g_{\text{eff}}=0$. 

We consider two different tunable-coupler-enabled leakage removal schemes: $\text{Coupler}_{20}$ and $\text{Pumped}_{21}$. The $\text{Coupler}_{20}$ scheme consists of the transmons and the tunable coupler configured as described above. This scheme takes advantage of the $\ket{20} \leftrightarrow \ket{02}$ transition, where the leakage propagates between the two transmons in roughly~\SI{1}{\micro\second}, see Fig.~\ref{fig:leakage_dynamic}(a). The $\text{Pumped}_{21}$ scheme takes advantage of the faster $\ket{21} \leftrightarrow \ket{12}$ transition, where the leakage propagates between the two transmons in roughly~\SI{20}{\nano\second}, see Fig.~\ref{fig:leakage_dynamic}(e). Since the dissipation on the second transmon also removes the single excitation that enables the faster leakage propagation, we repopulate the second transmon by repeatedly applying $X$ gates. In the simulations, we implement the $X$ gate by applying a Gaussian pulse with a gate duration of~\SI{14}{\nano\second} and~\SI{4}{\nano\second} pauses in between.

Lastly, we introduce a third passive leakage removal scheme, named as directly coupled scheme, which we use to benchmark the two previously introduced tunable-coupler-enabled leakage removal schemes. This scheme consists of two directly coupled (capacitive coupling) transmons tuned so that the transition $\ket{20} \leftrightarrow \ket{11}$  is resonant. As in the previous schemes, we add dissipation to the second transmon to remove the leakage from the system. Due to the resonance condition the transition frequencies of the two transmons are detuned so that $\Delta_{12} = U$, which offers some protection from the dissipation for the qubit single excitation level. We use this scheme as a baseline to determine whether tunable couplers provide an advantage over direct capacitive coupling in terms of coding qubit protection and leakage removal. A similar approach to the directly coupled scheme has previously been used to implement leakage $\textrm{SWAP}$ gates for leakage removal~\cite{miao_overcoming_2023, chen_fast_2024, yang_coupler-assisted_2024}. In total, we numerically simulate three different passive leakage removal schemes and plot the corresponding leakage and qubit population time dynamics, see Fig.~\ref{fig:lru_comparison}(a-d). 

The operation of this leakage removal unit can be roughly decomposed into two sequential processes, although both happen simultaneously: first, the coherent transfer of leakage population away from the coding transmon, followed by the removal of this population from the dissipative transmon. The first stage is governed by unitary leakage dynamics, determined by the oscillation amplitudes and frequencies detailed in the preceding sections. The second stage involves non-unitary population removal from the second site, and subsequently from the system entirely, via engineered dissipation. Given that these specific leakage removal units consist of only two sites, the characteristic timescales of the two underlying processes can be assumed to be equivalent. Importantly, the dissipation rate has an optimal value in relation to leakage removal~\cite{martinvazquez_2025}. We investigate this dissipation mechanism using both numerical and analytical methods. First, we simulate the three schemes with their respective optimal dissipation rate. For the two tunable-coupled schemes, these are $\Gamma/2\pi = $~\SI{0.4}{\mega\hertz} and $\Gamma/2\pi = $~\SI{9}{\mega\hertz}, for the $\text{Coupler}_{20}$ and $\text{Pumped}_{21}$, respectively. For the directly coupled scheme, we use $\Gamma/2\pi = $~\SI{20}{\mega\hertz}. Note that the dissipation rates used do not optimize the qubit lifetimes. 

Analytically, we evaluate the norm decay of the effective state within the framework of quantum trajectories evolved under a non-Hermitian effective Hamiltonian, taking this decay as a direct metric for excitation removal~\cite{martinvazquez_2025}. Consequently, the total excitation population decays as $e^{-t/T}$, where the characteristic decay time $T$ depends explicitly on the engineered dissipation rate $\Gamma$. We define $T_\star$ for leakage decay time and $T_1$ for qubit subspace excitation decay time. We analyze their behavior across the three distinct schemes described above. For the directly coupled scheme and $\text{Coupler}_{20}$ regimes, the analytical framework from Ref.~\cite{martinvazquez_2025} maps straightforwardly by accounting for the modified effective hopping strengths. For the $\text{Pumped}_{21}$ scheme, incorporating the pulsed $X$-gate drive complicates the analytical treatment. To circumvent this, we estimate an upper bound by assuming the drive continuously and perfectly maintains a single excitation at the second site, while the engineered dissipation removes only one excitation during a jump event, always leaving a single excitation remaining at the dissipative transmon. To account for the finite duration required to create the excitation, we scale the effective hopping strength by a factor $\epsilon \in (0, 1)$, mapping $J_{\text{eff}} \to \epsilon J_{\text{eff}}$. Within this parameterization, the limiting case $\epsilon = 1$ represents the upper bound of an idealized, infinitely fast $X$-gate, whereas for $\epsilon = 0$ there is no $X$-gate application and we should consider the model corresponding to the $\text{Coupler}_{20}$ scheme described before. This lower boundary establishes a strict operational requirement for the $\text{Pumped}_{21}$ scheme: the rate of the $X$-gate application must be faster than the effective hopping rates of the underlying $\text{Coupler}_{20}$ scheme. Although each effective hopping strength strictly involves a different scaling factor, we assume a uniform $\epsilon$ across all cases to preserve a tractable, generalized qualitative understanding.

We extract the optimal leakage removal rates from the different decay times. The directly coupled scheme, shares the algebraic structure of the high optimal rate regime corresponding to leakage disintegration described in Ref.~\cite{martinvazquez_2025}, although tuned to exact resonance. The characteristic decay time scales as $T_\star \approx (U^2 + \Gamma^2) / (8 J^2 \Gamma)$, which is minimized for $\Gamma \approx 4J$, resulting here in $\Gamma /2\pi \approx 20~\text{MHz}$. The $\text{Coupler}_{20}$ case, mirrors the low optimal rate regime associated with leakage propagation in Ref.~\onlinecite{martinvazquez_2025}, scaled by the effective coupler-mediated hopping strength $J_{20}$ found above. The resulting decay time is given by $T_\star \approx (2J_{20}^2 + \Gamma^2) / (2J_{20}^2 \Gamma)$, which is minimized for $\Gamma \approx \sqrt{2}|J_{20}|$, yielding $\Gamma / 2\pi \approx 0.7~\text{MHz}$ for the parameters considered here. This slight deviation arises from our perturbative approximations performed here, and the simplified treatment of the decay times. The exact dynamics are more complex because the optimal condition varies over time, for an explanation and exact equation see Ref.~\cite{martinvazquez_2025}. The $\text{Pumped}_{21}$ scheme leakage evolution can be modeled as an effective single-excitation dynamics $\ket{21} \leftrightarrow \ket{12} $, where we only remove one excitation from the second site when the leakage is there. The characteristic decay time is given by $T_\star \approx [2(\epsilon J_{21})^2 + \Gamma^2] / [2(\epsilon J_{21})^2 \Gamma]$, which is minimized when $\Gamma \approx \sqrt{2}|\epsilon J_{21}|$, resulting in $\Gamma/2\pi \approx 15\epsilon~\text{MHz}$ for the parameters considered here. Consequently, a lower $X$ gate pulse efficiency shifts the optimum rate to smaller values.

\subsection{Coupler in the ground state}
In Fig.~\ref{fig:lru_comparison}(a), we plot the population of the first transmon, the coding qubit, as a function of time for the three different leakage removal schemes. For the two tunable-coupler-enabled leakage removal schemes the initial state is $\ket{20_{\rm c}0}$. Similarly, for the two capacitively coupled transmons the initial state is $\ket{20}$. Note that both the capacitively coupled transmons and the repeatedly driven tunable coupler scheme remove single leakage excitations. Therefore, the state of the coding qubit after leakage removal is the first excited state. On the contrary, without the repeated driving, the two transmons in the tunable coupler scheme undergo the $\ket{20_{\rm c}0} \rightarrow \ket{00_{\rm c}2}$ transition, from which it follows that the state of the coding qubit after leakage removal is the ground state.

We find that the directly coupled transmons and the $\text{Pumped}_{21}$ scheme have comparable leakage removal times even though the dissipation rate used for the repeatedly driven scheme is roughly half of the dissipation rate used for the directly coupled transmons. This highlights the effectiveness of the $\text{Pumped}_{21}$ scheme in removing leakage. We find the leakage decay times to be roughly \SI{32}{\nano\second} in these two schemes. Out of the three, the $\text{Coupler}_{20}$ scheme performs the slowest, with leakage decay times of roughly~\SI{700}{\nano\second}. Analytical results are consistent with numerical simulations. Directly coupled and $\text{Pumped}_{21}$ schemes exhibit comparable leakage decay times, with $T_\star \approx 1/J \approx 32~\text{ns}$, and $T_\star \approx \sqrt{2}/|\epsilon J_{21}| \approx 21/\epsilon~\text{ns}$, respectively. In contrast, the $\text{Coupler}_{20}$ scheme yields a significantly larger leakage decay time $T_\star = \sqrt{2}/|J_{20}| \approx 0.45~\mu\text{s}$.

In Fig.~\ref{fig:lru_comparison}(c) we show the decay of the first excited state of the coding qubit under the different leakage removal schemes. For the tunable-coupler-enabled leakage removal schemes we use the initial state $\ket{10_{\rm c}0}$, and in the directly coupled transmons scheme the initial state $\ket{10}$. We find that the tunable coupler is very effective at protecting the coding qubit from the dissipation of the leakage removal unit. Both the $\text{Coupler}_{20}$ and the $\text{Pumped}_{21}$ scheme show minimal decay of the first excited state of the coding qubit. For the directly coupled transmons, we see that the detuning is not sufficient at protecting the qubit from the dissipation, leading to a $T_1$ time of roughly~\SI{28}{\micro\second}. Analytically, we find that $\text{Coupler}_{20}$ provides the strongest protection for the qubit. It correspond to one excitation undergoing resonant oscillations. Since the system operates under the strict condition where the qubit dynamics is suppressed $g_{\text{eff}} = 0$, the qubit decay time $T_1 \approx (8g_{\text{eff}}^2 + \Gamma^2) / (4g_{\text{eff}}^2 \Gamma)$, diverges toward infinity. In contrast, the directly coupled scheme, corresponds to one excitation oscillating off resonance by $\Delta_{12}$. The population within the qubit subspace decays with $T_1 \approx (4 \Delta_{12}^2 + \Gamma^2) / (4 J^2 \Gamma)$. The optimal rate occurs at $\Gamma \approx 2|\Delta_{12}|\sim 2\pi \times \SI{600}{MHz}$, which is far from the leakage removal rate. Evaluated at the optimal rate for leakage removal, the qubit population decays with $T_1 \approx J^{-1}[1 + \Delta_{12}^2 / (4J^2)] \approx 29~\mu\text{s}$. For the $\text{Pumped}_{21}$ case, the relevant dynamics follows the $\ket{11} \leftrightarrow \ket{02}$ transition, where the dissipation again eliminates only a single excitation from the second transmon when there leakage is there. The decay time scales as $T_1 \approx (U^2 + \Gamma^2) / [2(\epsilon J_{20,11})^2 \Gamma]$. Evaluated at the optimal leakage removal rate, the decay time results in $T_1 \approx 15/\epsilon^3~\mu\text{s}$. This cubic dependence demonstrates that even a minor reduction in the effective hopping strength caused by the $X$-gate drive provides a significant protecting effect for the computational qubit.

Then, the finite duration required to execute the $X$-gate controls a critical trade-off: it shifts the optimal parameter space for leakage removal and increases the associated decay times. Crucially, while this scaling incurs only a minor penalty for leakage removal $\mathcal{O}(1/\epsilon)$, it provides a major advantage by cubically suppressing computational qubit decay $\mathcal{O}(1/\epsilon^3)$. For example, for the value found numerically for the optimal rate $\Gamma/2\pi = 9~\text{MHz}$, we have $\epsilon =0.6$, so that the leakage decay time results in $T_\star \approx 35 ~\text{ns}$ being the same as the direct coupler case and the qubit decay time in $T_1 \approx 70 ~\mu\text{s}$ being larger than the directly coupled scheme [a reduction in the qubit population for the $\text{Pumped}_{21}$ scheme can be observed in the inset of Fig.~\ref{fig:lru_comparison}(c)]. From the numerical analysis, we find $T_1 \approx 1000~\mu\text{s}$; the variance relative to the analytical model arises from our idealized assumptions and the complex combinations of dynamics triggered by the $X$-gate pulses. 

\subsection{Coupler with one excitation}
In tunable coupler setups, leakage can enter the coupler for example from two qubit gates~\cite{sung_2021} or stray coupling~\cite{yang_coupler-assisted_2024}, or from the weak interaction between the qubit and the coupler, visible in Fig.~\ref{fig:leakage_dynamic}(a) and (e) as small-amplitude rapid oscillations. Since tunable couplers are designed to operate in their ground state, leakage in the coupler can compromise their intended function. Thus, an effective leakage removal unit should be able to remove the leakage in the coupler. We plot the population of the coupler for an initial state $\ket{01_{\rm c}0}$ in Fig.~\ref{fig:lru_comparison}(b), for the two tunable-coupler-enabled leakage removal schemes. We find that only the $\text{Pumped}_{21}$ scheme is able to remove leakage in the coupler in a reasonable time, with a coupler leakage decay time of~\SI{320}{\nano\second}. We can have an analytical understanding of this by returning to the bare system Hamiltonian of Eq. \eqref{H_full}, and analyze the population transfer directly from the coupler to the dissipative transmon. The leakage decay time of $\text{Coupler}_{20}$ scales as $T_\star \approx (4\Delta_{c}^2 + \Gamma^2) / (4g_c^2 \Gamma)$. The optimal removal rate for this dynamics would require a high dissipation of $\Gamma \approx 2|\Delta_c| \approx 2\pi \times$~\SI{990}{\mega\hertz}; for the parameters considered here the decay time is $T_\star \approx 22~\mu\text{s}$. For the $\text{Pumped}_{21}$ scheme, we have that $T_\star \approx [(\Delta_{c} - U)^2 + \Gamma^2] / [4(\epsilon g_c)^2 \Gamma]$. At the optimal operating rate for leakage removal, this dynamics has a decay time of $T_\star \approx 40/\epsilon^3~\text{ns}$, considerable smaller than the $\text{Coupler}_{20}$ scheme where even with the correction of $\epsilon=0.6$, we get $T_\star\approx 185~\text{ns}$ being still two orders of magnitude smaller.

Since the leakage effectively changes the operating point of the tunable coupler we find in Fig.~\ref{fig:lru_comparison}(d) that leakage in the coupler compromises the tunable couplers ability to protect the coding qubit. We can make a rough analytical study by considering that the dissipative removal of excitations have two different time scales: first the excitation from the first transmon hops to the second transmon with a modified hopping strength due to the populated coupler and is removed [Fig.~\ref{fig:lru_comparison}(d)], and second, the excitation from the coupler hops to the second transmon and is removed as described previously in Fig.~\ref{fig:lru_comparison}(b). For the $\text{Coupler}_{20}$ scheme, we find that the modified effective hopping strength shifts to $g_{\text{eff}}' = g_q - 2g_c^2 / (\Delta_{c} + U) \approx -1.3~\text{MHz}$, as derived in Eq.~\eqref{H_eff}. This yields a qubit decay time of $T_1 \approx [8(g_{\text{eff}}')^2 + \Gamma^2] / [4(g_{\text{eff}}')^2 \Gamma]$, with an optimal removal rate of $\Gamma \approx 2\sqrt{2}|g_{\text{eff}}'|$. Since we are operating near this condition to remove the leakage, this yields a decay time of $T_1 \approx 0.48~\mu\text{s}$ for the qubit population. For the $\text{Pumped}_{21}$ scheme, the effective hopping strength increases slightly to $g_{\text{eff}}'' \approx - 7\epsilon~\text{MHz} $, leading to a qubit decay time of $T_1 \approx 10/\epsilon^3~\mu\text{s}$, or $T_1 \approx 46~\mu\text{s}$ considering $\epsilon=0.6$.

\begin{figure*}
  \centering
  \includegraphics[width=0.85\linewidth]{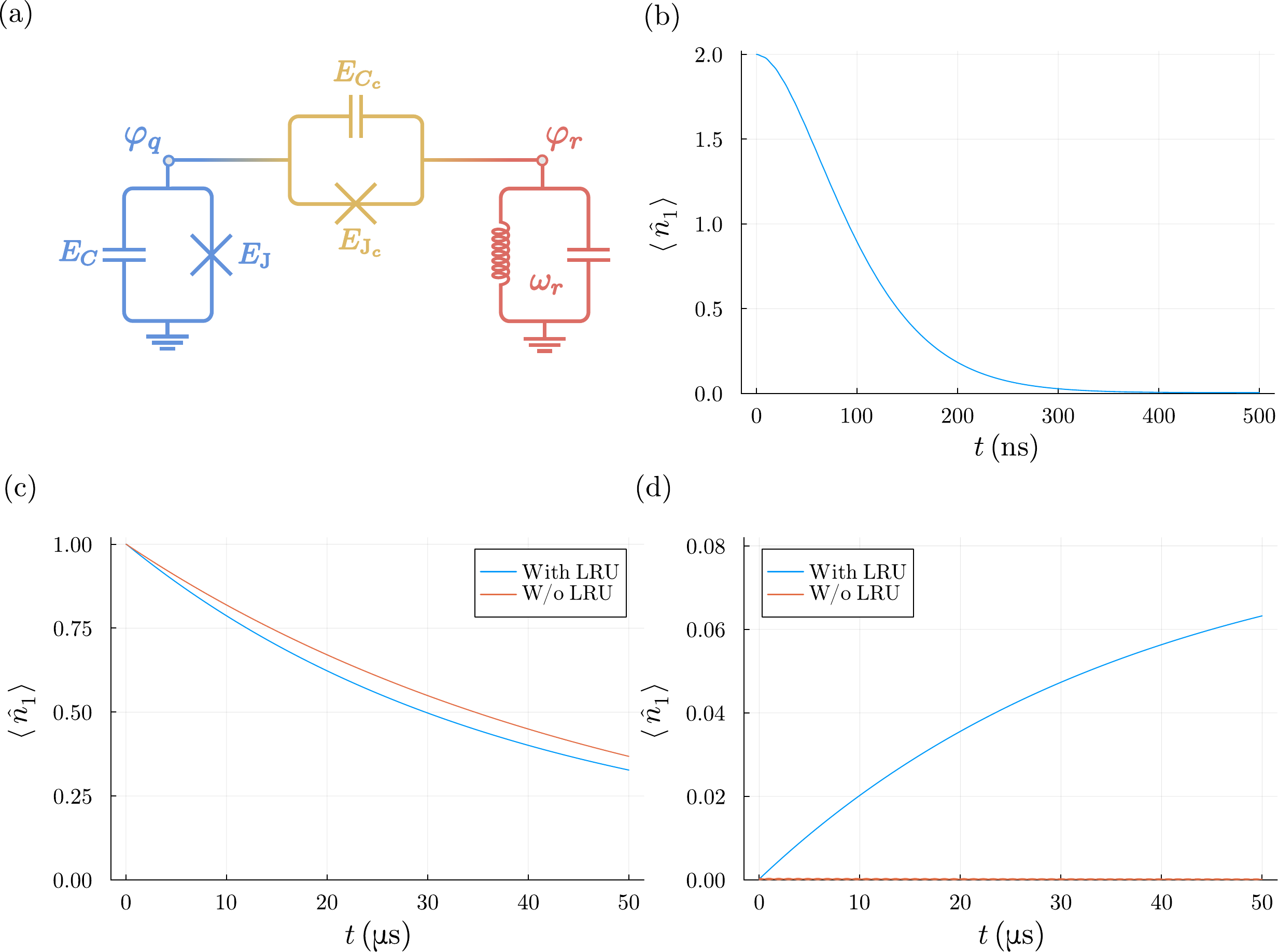}
  \caption{\label{fig:junction_readout} (a) Circuit schematic of the junction readout scheme. (b) Leakage dynamics for the initial state $\ket{2_q0_r}$. (c) Qubit dynamics for the initial state $\ket{1_q0_r}$ where the qubit has an intrinsic decay time of $T_1 = \SI{50}{\micro \second}$. We characterize the effect of the leakage removal on the qubit by comparing dynamics with (orange) and without (blue) the leakage removal dissipation.(d)~Effect of the leakage removal on the ground state of the qubit $\ket{0_q0_r}$ following from the two-mode squeezing and the leakage removal dissipation on the resonator.}
\end{figure*}

\section{Passive Leakage Removal With Junction Readout }\label{sec:LRU2}

We have established that tunable couplers can efficiently control the propagation of leakage while keeping the computational states of the qubit protected, making it a promising candidate for realizing passive leakage removal units (LRU)s. In a practical QPU, however, this approach requires either repurposing some qubits to act as LRUs, reducing the number of computational qubits, or introducing additional components, increasing the per-qubit footprint. Furthermore, a qubit assigned to the role of leakage removal unit requires a source of dissipation, such as an engineered environment for qubit reset or a lossy resonator used for dispersive readout~\cite{blais_circuit_2021}.

Besides the dispersive readout scheme, a nonlinear readout scheme known as junction readout has recently emerged as an alternative, offering fast and high-fidelity readout~\cite{chapple_balanced_2025, beaulieu_fast_2026, wang_longitudinal_2025}. Unlike dispersive readout, junction readout relies on a nonperturbative cross-Kerr interaction between the qubit and the readout resonator, providing improved protection against Purcell decay~\cite{blais_circuit_2021} and measurement-induced state transitions~\cite{sank_measurement-induced_2016,shillito_dynamics_2022, cohen_reminiscence_2023, khezri_measurement-induced_2023, dumas_measurement-induced_2024}. Here we propose that the junction-readout setup can serve a secondary role as a passive leakage removal unit, reusing the existing readout infrastructure to supply the much-needed dissipation source.

In junction readout, the transmon qubit and the readout resonator are coupled through a Josephson junction and a linear capacitor in parallel, as illustrated in Fig.~\ref{fig:junction_readout}(a). The junction produces the nonlinear interactions used for both readout and leakage removal, while the capacitor is essential for protecting the computational states by canceling out the single-excitation exchange interaction. This cancellation mechanism is similar to the destructively interfering paths of the tunable-coupler setup, but without requiring any additional mediator modes. We describe the system with the same Hamiltonian as in Ref.~\cite{chapple_balanced_2025}. However, to build better intuition, we separate the single-mode contributions of the coupling junction from the interaction and split the Hamiltonian into three parts, $\hat{H}_\text{readout} = \hat H_q + \hat H_r + \hat H_\text{int}$, corresponding to the transmon, the resonator and their interaction, with definitions
\begin{align}
    \hat{H}_q &= 4 E_C \hat n_q^2 - \tilde E_\text{J} \cos \hat \varphi_q \label{H_q_junction}, \\
    \hat{H}_r &= \hbar \omega_r \hat a^\dagger \hat a - E_{\text{J}_c} \cos \hat \varphi_r \label{H_r_junction}, \\
    \hat{H}_\text{int} &= 8 E_{C_c} \hat n_q \hat n_r + E_{\text{J}_c} \cos \hat \varphi_q + E_{\text{J}_c} \cos \hat \varphi_r \notag \\  
    & \ \ \ - E_{\text{J}_c} \! \left( \cos \hat \varphi_q \cos \hat \varphi_r + \sin \hat \varphi_q \sin \hat \varphi_r\right).
    \label{H_int_junction}
\end{align}
Here, $\hat \varphi_q$ ($\hat \varphi_r$) and $\hat n_q$ ($\hat n_r$) denote the phase and charge number operators of the transmon (resonator), satisfying the commutation relation $\left[ \hat \varphi_k, \hat n_k \right] = \rm{i}$ with $k \in \{ q,r \}$. The Josephson and charging energies of the transmon are $E_\text{J}$ and $E_C$, while $E_{\text{J}_c}$ and $E_{C_c}$ denote the Josephson and charging energies of the coupler. The single-mode contribution of the coupling junction adds to the bare transmon potential, renormalizing its Josephson energy to $\tilde E_\text{J} = E_\text{J} + E_{\text{J}_c}$ and thereby shifting the transmon frequency. The bare resonator frequency is $\omega_r$, and the resonator operators are written in ladder form as $\hat\varphi_r = \varphi_\text{zpf,r}(\hat a + \hat a^\dagger)$ and $\hat n_r = \mathrm{i}(\hat a^\dagger - \hat a)/(2\varphi_\text{zpf,r})$, where $\varphi_\text{zpf,r}~=~(2\pi/\Phi_0)\sqrt{\hbar Z_r/2}$ is the amplitude of the phase zero-point fluctuations and $Z_r$ denotes the resonator characteristic impedance.

The interaction of interest in the junction-readout setup is the $\cos\hat\varphi_q\cos\hat\varphi_r$ term. In particular, its fourth-order expansion term, $\hat\varphi_q^2\hat\varphi_r^2$, is responsible for both the cross-Kerr interaction used for readout and the two-excitation exchange between the states $\ket{2_q0_r}$ and $\ket{0_q2_r}$ that we exploit for leakage removal. Here, the notation $\ket{i_q j_r}$ denotes the product states formed from the eigenstates of the Hamiltonians in Eqs.~\eqref{H_q_junction} and~\eqref{H_r_junction}.

While passive leakage removal requires the transition $\ket{2_q0_r} \leftrightarrow \ket{0_q2_r}$ to be resonant, the Hamiltonian is simultaneously tuned so that the single-excitation exchange through the capacitive and inductive paths interferes destructively, imposing the condition
\begin{align}
     \braket{1_q 0_r|\hat n_q \hat n_r|0_q 1_r} \! - \! \frac{E_{\text{J}_c}}{8 E_{C_c}} \! \braket{1_q 0_r|\sin \hat \varphi_q \sin \hat \varphi_r|0_q 1_r} \! = \! 0.
    \label{single_photon_condition}
\end{align}
The importance of suppressing the $\ket{1_q0_r}\leftrightarrow\ket{0_q1_r}$ transition is further highlighted by the resonance requirement for the $\ket{2_q0_r}\leftrightarrow\ket{0_q2_r}$ transition, leaving only a small detuning $\Delta_{qr}=-(U_q-U_r)/2$ between the single-excitation states, where $U_q$ and $U_r$ denote the anharmonicities of the two modes. The single-excitation states are therefore too close for detuning alone to suppress their exchange, making the interference condition essential. 

However, realizing effective two-excitation exchange simultaneously with single-excitation exchange cancelation does not come without a cost. The price of this cancelation is that the same two paths interfere constructively for the two-mode squeezing interaction, which couples the computational states out of their subspace, generating nonzero amplitudes for the transitions $\ket{0_q0_r}\leftrightarrow\ket{1_q1_r}$ and $\ket{1_q0_r}\leftrightarrow\ket{2_q1_r}$. Although these transitions are heavily suppressed by the large frequency detuning, the intrinsic dissipation of the readout resonator opens a channel for Purcell decay that affects the qubit lifetime~\cite{beaulieu_fast_2026}.

In Fig.~\ref{fig:junction_readout}, we provide an example of a junction-readout setup used for passive leakage removal.
The numerical simulations are performed with the full Hamiltonian of Eqs.~\eqref{H_q_junction}-\eqref{H_int_junction}, using circuit parameters $E_\text{J}/h =$~\SI{20.0}{\giga\hertz} and $E_C/h =$~\SI{200}{\mega\hertz} for the qubit, $E_{\text{J}_c}/h =$~\SI{1.0}{\giga\hertz} and $E_{C_c}/h =$~\SI{1.7}{\mega\hertz} for the coupler, and $\omega_r/2\pi =$~\SI{5.4}{\giga\hertz} and $Z_r =$~\SI{55.9}{\ohm} for the resonator.
These parameters enforce resonant two-excitation exchange and suppress single-excitation exchange while keeping a realistic transmon transition frequency
$\omega_{01}^{(q)}/2\pi =$~\SI{5.582}{\giga\hertz}
and anharmonicity 
$U_q/2\pi =$~\SI{219}{\mega\hertz}.
The readout resonator has a transition frequency
$\omega_{01}^{(r)}/2\pi =$~\SI{5.480}{\giga\hertz}
and acquires a small anharmonicity
$U_r/2\pi =$~\SI{0.36}{\mega\hertz}
from the nonlinearity of the coupling junction. This residual anharmonicity does not affect the leakage-removal function but it does matter for readout, and is therefore designed to be small. For these parameters we find a coupling strength $g_{20,02}/2\pi = \braket{0_q2_r|\hat H_\text{int}|2_q0_r}/h =$~\SI{-1.7}{\mega\hertz}, corresponding to a SWAP time of roughly \SI{140}{\nano\second}. We find the leakage removal to be efficient by choosing $\kappa/2\pi = 3.4\,\text{MHz}$, which results in a leakage decay time of $T_\star \approx \SI{125}{\nano\second}$~[Fig.~\ref{fig:junction_readout}(b)].

The junction-readout setup is protected from Purcell decay through the single-excitation exchange, but not from Purcell-induced effects through the two-mode squeezing. As shown in Figs.~\ref{fig:junction_readout}(c)--(d), resonator dissipation causes weak coherent oscillations between $\ket{0_q0_r}$ and $\ket{1_q1_r}$ to slowly accumulate population in $\ket{1_q0_r}$. A similar effect takes place between the states $\ket{1_q0_r}$ and $\ket{2_q1_r}$, slowly accumulating population in the leakage state $\ket{2_q0_r}$. The matrix elements of these transitions, $g_{00,11}/2\pi = -114\,\text{MHz}$ and $g_{10,21}/2\pi = -161\,\text{MHz}$, are relatively large, but the transitions are significantly weakened by their detunings that exceed 10 GHz. The resulting Purcell-induced transition times can be estimated as $T_\text{1,Purcell}^{-1} = 2 \pi \kappa\,(g/\Delta)^2$, where $g$ and $\Delta$ are the coupling strength and detuning of the transition. For the two-mode-squeezing transitions $\ket{0_q0_r}\leftrightarrow\ket{1_q1_r}$ and $\ket{1_q0_r}\leftrightarrow\ket{2_q1_r}$, this gives Purcell-induced transition timescales of roughly \SI{440}{\micro\second} and \SI{215}{\micro\second}, respectively. These are comparable to the upper end of typical transmon lifetimes, so the two-mode squeezing does not introduce significant additional errors.

Although the two-excitation exchange interaction already achieves removal times comparable to other passive schemes~\cite{thorbeck24, martinvazquez_2025}, we expect further improvements to be possible. For instance, a photon-enhanced single-excitation exchange between the resonantly tuned states $\ket{2_q0_r}$ and $\ket{1_q1_r}$ yields a matrix element more than three times larger when compared with the two-excitation exchange, potentially enabling considerably faster removal with only minor changes to the parameters.

The parameters used for the example in Fig.~\ref{fig:junction_readout} were chosen with the readout function in mind as well. We note that the operating point used for leakage removal is not ideal for readout: the resonant two-excitation exchange can disturb the qubit state when the resonator is driven to a coherent state during measurement. This can be avoided by introducing a large separation between the transmon and resonator frequencies before measuring the qubit. For instance, replacing the Josephson junctions of the transmon and the coupler with tunable SQUIDs would allow the frequencies to be detuned while keeping the single-excitation exchange suppressed. For example, tuning the Josephson energies of the qubit and the coupler to $E_\text{J}/h =$~\SI{12.5}{\giga\hertz} and $E_{\text{J}_c}/h =$~\SI{0.795}{\giga\hertz} brings the qubit frequency down to $\omega_{01}^{(q)}/2\pi =$~\SI{4.407}{\giga\hertz}, suppressing the two-excitation exchange while the single-excitation exchange remains canceled by the destructive interference. At these parameters, the cross-Kerr interaction between the qubit and resonator modes is approximately $\chi/2\pi =$~\SI{3.6}{\mega\hertz}. We note, however, that these parameters illustrate how the setup is switched between the leakage-removal and readout functions and are not optimized for readout performance.

\section{Discussion and conclusions}\label{sec:disc}
In this work, our focus was in mobility of leakage excitations in transmons with tunable couplers. First, we found that coupler operating points that cancel either single-excitation exchange or $\textrm{ZZ}$ interactions do not simultaneously suppress leakage excitation hopping originating in the nonlinearity of transmons and couplers. By utilizing Schrieffer-Wolf-type perturbation theory, we derived effective analytical models for the leakage excitation dynamics and confirmed them with numerical methods. In a resonant case, we find leakage hopping rates in the range of ~\SIrange{0.8}{10}{\mega\hertz}. In general, the coupler operating point that cancels $\textrm{ZZ}$ interaction produces much smaller leakage hopping strengths.  Leakage can tunnel to the next-nearest neighbor with hopping rates of approximately~\SI{0.1}{\mega\hertz} in a scenario where the nearest site is far-detuned but the next-nearest is in resonance for the leakage excitations.

In superconducting quantum processors, leakage excitations degrade the fidelity of quantum gates and reduce effectiveness of quantum error correction approaches. Our results on leakage mobility in transmons with tunable couplers can be applied to either localize or mobilize leakage excitations depending as needed. Localization could be a desirable feature if one wants to prevent traveling leakage excitations from inducing stochastic gate errors, or to remove leakage through, e.g., dynamic switching of data and auxiliary qubits. A convenient way to realize localization is via frequency detuning beyond the magnitude of nearest and next-nearest leakage hopping rates. In 2D arrays, one transmon has \SI{3}{} or \SI{8}{} next-nearest neighbors in heavy-hexagonal or square-grid qubit topologies and each of them should sufficiently detuned. Conversely, mobile leakage excitations could be useful for creating effective leakage removal units, transporting leakage excitations to distant leakage removal units, or creating data-auxiliary qubit exchange~\cite{Camps24} through the leakage dynamics itself. 

We proposed two main leakage removal units based on mobilized leakage excitations. The first one directly applies the tunable transmon couplers, achieving leakage removal times of approximately~\SI{30}{\nano\second}, while the second one employs a junction readout circuit~\cite{chapple_balanced_2025, beaulieu_fast_2026, wang_longitudinal_2025} with leakage removal times of approximately~\SI{125}{\nano\second} for realistic and experimentally relevant parameters. These times translate to a $99\%$ removal of the leaked population within approximately $157~\text{ns}$ and $300~\text{ns}$, respectively, without inducing any relevant perturbation to the underlying computational subspace. Additionally, we introduced the $\text{Coupler}_{20}$ scheme. Although it exhibits a relatively slow characteristic leakage decay time of $700$~ns, implying $99\%$ population removal in $3.2~\mu\text{s}$, it protects the computational subspace excitations perfectly, provided that the $g_{\text{eff}} = 0$ condition holds.

Crucially, these architectures function as passive, always-on leakage removal units~\cite{thorbeck24, martinvazquez_2025}. By operating continuously in the background, they circumvent the need for event-triggered control pulses, allowing them to run seamlessly during simultaneous gate execution and quantum error correction cycles; the $\text{Pumped}_{21}$ configuration requires only the continuous application of a baseline driving field. In contrast, traditional active removal techniques require targeted control pulses on the coding qubit, state-swap operations, or feedback-driven correction cycles~\cite{bultink2020, varbanov20, McEwen2021, miao_overcoming_2023, Marques23}. 

The schemes proposed here mark a substantial advancement in the speed of removal over our previously designed passive leakage removal unit, which reported a total leakage decay time of $T_\star \approx 2.2~\mu\text{s}$ and a removal time from the coding sites of $\sim 1.25~\mu\text{s}$~\cite{martinvazquez_2025}. Furthermore, the proposed schemes match the performance of alternative passive strategies, such as configurations introducing custom high-order band-pass filters that achieve comparable leakage decay times of $33~\text{ns}$~\cite{thorbeck24}. Importantly, although active methods are typically faster than passive alternatives, our new proposed designs operate on similar timescales and even surpass several active protocols. For context, representative active methods report an $80\%$ leakage reduction in $\approx 1~\mu\text{s}$~\cite{miao_overcoming_2023}, a $99\%$ reduction in $\approx 220~\text{ns}$~\cite{Marques23}, swap-based evacuation in $\approx 250~\text{ns}$~\cite{McEwen2021}, and feedback-assisted schemes integrated with quantum error correction in $\approx 1~\mu\text{s}$~\cite{bultink2020, varbanov20}. More recently, an active leakage removal unit achieved a $98\%$ removal efficiency in $500~\text{ns}$ in the context of error correction~\cite{xin2025}, while a fast flux-pulse protocol utilizing a transmon coupled to a readout resonator achieved a $99\%$ removal efficiency in $50~\text{ns}$~\cite{lacroix2025}. Our passive schemes achieve similar or superior performance thresholds without incurring the significant control-layer overhead typical of active methods. However, our schemes have been designed for leakage on the level $\ket{2}$, and higher levels would require different settings.

The dissipation required by the $\text{Coupler}_{20}$ and $\text{Pumped}_{21}$ leakage removal schemes can be engineered by taking advantage of the Purcell effect~\cite{reed_fast_2010}. Nevertheless, a transmon that is simultaneously strongly dissipative and resonant with a neighboring transmon is not typically available in current transmon-based quantum processors. An alternative approach is to replace the passive dissipation with the readily available qubit reset operation of the QPU. The leakage dynamics shown in Fig.~\ref{fig:leakage_dynamic}(a) and (e) can be used to realize a leakage $\textrm{SWAP}$ gate, followed by qubit reset. In the $\text{Pumped}_{21}$ scheme the leakage $\textrm{SWAP}$ gate is realized by applying the $X$ gate, followed by a \SI{20}{\nano\second} idle. Since the typical qubit reset protocols have reset times of a few hundred nanoseconds~\cite{zhou_rapid_2021}, they add considerable overhead to the leakage removal schemes presented here. Note that, since the reset is performed from the perspective of the algorithm on an auxiliary qubit, it does not compromise the passive nature of the leakage removal schemes.

Looking forward, we expect that the results and methods presented can be further applied and developed for other tunable couplers and qubits. We studied here passive mobility of leakage excitations in a superconducting quantum processor. To get a full picture of how leakage affects the performance of quantum computing algorithms or error-correction protocols, one needs to add the influence of quantum gates and measurements, as well as the presence of other qubits and couplers, which makes the situation resemble that of many-body dynamics.

\section*{Acknowledgments}
We thank Sami Laine, Tuure Orell and Jani Tuorila for useful discussions. We acknowledge funding by Kvantum Institute of the University of Oulu, and the Research Council of Finland under Grants Nos. H2Future Profi7 352788 and 355824. This publication is part of the research and applied innovation project Quantum mechanics and biology: new theoretical frameworks and analytictechnical applications SOL2024-31572, co-financed by the “UE - Ministerio de Hacienda y Función Pública - Fondos Europeos - Junta de Andalucía – Consejería de Universidad, Investigación e Innovación”.

\appendix
\section{Two transmons and a tunable couplers}
\label{App:couplers}
To derive the effective Hamiltonian within the transmon subspace, we employ a Schrieffer-Wolff transformation $\hat{H}_{\text{eff}} = e^{\hat{S}}\hat{H}e^{-\hat{S}}$~\cite{bravyi_2011}, where the anti-Hermitian generator $\hat{S}$ is defined by the commutation relation $[\hat{S}, \hat{H}_0] = -\hat{V}_c$, where $\hat{V}_c$ include all the hoping terms between transmons and the tunable coupler~\cite{yan_2018}. Because our analysis extends to excitations outside the computational qubit subspace, we utilize a generalized operator $\hat{S}$ to account for transitions involving higher-level transmon states. 
\begin{align}
    \hat{S}&=\sum_{j=1,2} \left( \hat{a}_j^\dagger\hat{a}^{}_c \frac{g_{jc}}{\Delta_{jc}-\hat{n}_jU_j+(\hat{n}_c-1)U_c} -  \text{H.c.}\right).
    \label{SW_operator}
\end{align}
The resulting effective Hamiltonian is subsequently projected onto the transmon manifold of interest. Throughout the expansion, we keep track of the perturbative order in terms of the parameter $\lambda \equiv |g_{jc}/\Delta_{jc}|$. When considering leakage states, we treat terms such as $g_c/U$ and $g_c/(\Delta_{jc} \pm U)$ as scaling with $\mathcal{O}(\lambda)$. We assume that the direct transmons coupling is of the order $|g_{12}| \sim \mathcal{O}(\lambda g_{jc})$. Note that across the paper, we consider that the coupler is in the ground state $n_c=0$, except in Fig.~\ref{fig:lru_comparison}(b),(d) from Sec.~\ref{Sec:lru}.

A critical consideration for this perturbative analysis is that the derived results remain valid only in the dispersive regime, well away from any resonance conditions. The presence of terms proportional to $\Delta_{jc}^{-1}$ and $(\Delta_{jc} \pm U)^{-1}$ implies that the analysis becomes singular and physically invalid in the vicinity of $\Delta_{jc} = 0$ and $\Delta_{jc} = \mp U$, respectively. Consequently, the accuracy of our effective Hamiltonian improves as the system is further detuned from these resonant points.
In the main text we only consider the case where the coupler frequency is lower than the qubit frequencies. Note that in this case the relevant detunings are positive ($\Delta_{jc} > 0$), and the pervasive perturbative term $g_c/(\Delta_{jc} - U)$ limits the accuracy of the fourth-order expansion due to a smaller denominator. Conversely, when the coupler frequency is tuned higher than the qubit frequencies, the negative detuning ($\Delta_{jc} < 0$) ensures that the fourth-order perturbative analysis remains accurate in most cases. In this alternative regime, the counterpart terms containing $g_c/(\Delta_{jc} + U)$ appear less frequently in the expansion, as they arise exclusively from higher-energy virtual leakage states localized within the coupler. 

While the main text provides the general second-order result, this Appendix presents the explicit expressions for the specific cases under study. In all cases, we explicitly derive perturbative two-dimensional systems where we study oscillations between two transmons. When the two transmons are on resonance, the effective hopping strength $J_{\text{eff}}$ directly determines the characteristic oscillation frequency, implying that suppressing the dynamics requires satisfying the condition $J_{\text{eff}} = 0$. When the transmons are off-resonance by an energy detuning $\Delta E$, the resulting dynamics are governed by the amplitude $4J_{\text{eff}}^2 / (\Delta E^2 + 4J_{\text{eff}}^2)$ and a generalized Rabi frequency $\sqrt{\Delta E^2 + 4J_{\text{eff}}^2}$. These analytical expressions are utilized throughout the text to analyze the role of frequency detuning across the various dynamical regimes. 

\subsection{$\ket{10} \leftrightarrow \ket{01}$}
First, we study the simplest case of one excitation oscillating between two transmons. The resulting effective Hamiltonian is
\begin{align}
    \frac{\hat{H}_{\text{eff}}}{\hbar}
    =
    \begin{pmatrix}
         \omega_1+E_{10}^{(2)}/\hbar &  g_{12}+ J^{(2)}+ J^{(4)} \\
        g_{12}+J^{(2)}+J^{(4)} & \omega_2+E_{01}^{(2)}/\hbar
    \end{pmatrix}
\end{align}
where $E_{10}^{(2)}=\hbar g_{1c}^2/\Delta_{1c}$ and $E_{01}^{(2)}=\hbar g_{2c}^2/\Delta_{2c}$, and the hopping terms are
\begin{align}
    J^{(2)}=&\frac{g_{1c}g_{2c}}{2}\left( \frac{1}{\Delta_{1c}}+\frac{1}{\Delta_{2c}}\right),\\
    J^{(4)}=&-\frac{1}{2} g_{12} \left(\frac{g_{1c}^2}{\Delta_{1c}^2}+\frac{g_{2c}^2}{\Delta_{2c}^2}\right) \nonumber \\
    &\ -\frac{1}{8}g_{1c}g_{2c}\left[\frac{g_{1c}^2}{\Delta_{1c}^2}  \left(\frac{1}{\Delta_{1c}}+\frac{7}{\Delta_{2c}}\right)\right. \nonumber \\
    & \qquad \qquad \quad +\left.\frac{g_{2c}^2}{\Delta_{2c}^2}  \left(\frac{1}{\Delta_{2c}}+\frac{7}{\Delta_{1c}}\right)\right].
\end{align}
The effective hopping strength is given by $g_{\text{eff}}=g_{12}+J^{(2)}+J^{(4)}$. 

\subsection{$\ket{20} \leftrightarrow \ket{02}$}
Second, we consider the case where there is a leakage in the first transmons and the second transmon is in the ground state. We obtain a three-dimensional effective Hamiltonian
\begin{align}
    &\frac{\hat{H}_{\text{eff}}}{\hbar}=\notag\\
    &\begin{pmatrix}
        \frac{E_{20}^{(0)}}{\hbar}+\frac{E_{20}^{(2)}}{\hbar} & \sqrt{2} g_{12}+J^{(2)}_1 & J^{(4)} \\
        \sqrt{2} g_{12}+J^{(2)}_1 & \frac{E_{11}^{(0)}}{\hbar}+\frac{E_{11}^{(2)}}{\hbar} & \sqrt{2} g_{12}+J^{(2)}_2 \\
        J^{(4)} & \sqrt{2} g_{12}+J^{(2)}_2 & \frac{E_{02}^{(0)}}{\hbar}+\frac{E_{02}^{(2)}}{\hbar}
    \end{pmatrix},
    \label{H_eff_20_11_02}
\end{align}
where $E_{20}^{(0)}/\hbar=2\omega_1-U_1$, $E_{11}^{(0)}/\hbar =\omega_1+\omega_2$, and $E_{02}^{(0)}/\hbar=2\omega_2-U_2$. The second order corrections to the energy are
\begin{align}
    E_{20}^{(2)}=&\hbar g_{1c}^2\left( \frac{2}{\Delta_{1c}-U_1}\right), \\
    E_{11}^{(2)}=&\hbar \left(\frac{g_{1c}^2}{\Delta_{1c}}+\frac{g_{2c}^2}{\Delta_{2c}}\right), \\
    E_{02}^{(2)}=&\hbar g_{2c}^2\left( \frac{2}{\Delta_{2c}-U_2}\right),
\end{align}
and the second order corrections to the hopping terms are
\begin{align}
    J^{(2)}_1=&\sqrt{2}\frac{g_{1c}g_{2c}}{2} \left(\frac{1}{\Delta_{1c}-U_1} +\frac{1}{\Delta_{2c}}\right), \\
    J^{(2)}_2=&\sqrt{2}\frac{g_{1c}g_{2c}}{2} \left(\frac{1}{\Delta_{1c}} +\frac{1}{\Delta_{2c}-U_2}\right).
\end{align}
The fourth order correction is given by
\begin{widetext}
\begin{align}
    J^{(4)}=&g_{12}g_{1c}g_{2c}\left[ \frac{2}{(\Delta_{2c}-U_2)(\Delta_{1c}-U_1)} -\frac{1}{\Delta_{2c}(\Delta_{1c}-U_1)}-\frac{1}{\Delta_{1c}(\Delta_{2c}-U_2)}\right] \nonumber \\
     &-\frac{g_{1c}^2 g_{2c}^2}{4}
     \left[ 
     \left(\frac{1}{ \Delta_{1c}\Delta_{2c}}+\frac{2}{(\Delta_{1c}+U_c)(\Delta_{2c}+U_c)} \right)\left( \frac{1}{(\Delta_{1c}-U_1)}+\frac{1}{(\Delta_{2c}-U_2)}\right)\right. \nonumber \\    
     & \qquad \qquad \quad \left. +\frac{3}{(\Delta_{1c}-U_1)(\Delta_{2c}-U_2)} \left( \frac{1}{\Delta_{1c}}+\frac{1}{\Delta_{2c}}-\frac{2}{(\Delta_{1c}+U_c)}-\frac{2}{(\Delta_{2c}+U_c)}\right)
     \right].
     \label{J_20_4th}
\end{align}
\end{widetext}
Note that in this case we have two dynamics with different time scales, but since the state $\ket{11}$ is quite detuned with respect to $\ket{20}$ and $\ket{02}$, we can understand the dynamics of the leakage oscillating as a single particle. Applying again a Schrieffer-Wolff transformation, we obtain a two-dimensional system in the $\ket{20}, \ket{02}$ space, with the matrix elements
\begin{align}
    J_{20}=&J^{(4)}\notag\\+&\hbar\left( \sqrt{2}g_{12}+J^{(2)}_1\right)\left( \sqrt{2}g_{12}+J^{(2)}_2\right)\Delta E^{-1},\\
    \frac{E'_{20}}{\hbar}=&\frac{E_{20}}{\hbar} +\frac{\hbar \left(\sqrt{2}g_{12}+J^{(2)}_1\right)^2}{E_{20}^{(0)}-E_{11}^{(0)}} \\
    \frac{E'_{02}}{\hbar}=&\frac{E_{02}}{\hbar}+\frac{\left(\hbar \sqrt{2}g_{12}+J^{(2)}_2\right)^2}{E_{02}^{(0)}-E_{11}^{(0)}}
\end{align}
where $\Delta E^{-1}=[(E_{20}^{(0)}-E_{11}^{(0)})^{-1}+(E_{02}^{(0)}-E_{11}^{(0)})^{-1}]/2$. When both transmons are on resonance, we have that $J^{(2)}=J^{(2)}_1=J^{(2)}_2 $, so the effective hopping is given by
\begin{equation}
    J_{20}=-\frac{\left( \sqrt{2}g_{12}+J^{(2)}\right)^2}{U}+J^{(4)}.
    \label{J_200}
\end{equation}
Note that here, the relevant dynamics $\ket{20} \leftrightarrow \ket{02}$ occurs at fourth order, and the terms $(\Delta_{jc}+U_c)^{-1}$ limit the range of validity also for negative detunings. To obtain a straightforward result, we calculate the amplitude probability for the transition of a leakage state from site $i$ to site $j$ by applying standard non-degenerate perturbation theory to the bare Hamiltonian~\eqref{H_full}. This calculation yields 
\begin{align}
    &\braket{2_i|\hat{H}|2_j}\notag\\
    &\quad=\frac{2g_{ij}}{\Delta_{ij}-U_i} +\frac{2g_{ij}g_{ic}g_{jc}}{(\Delta_{ij}-U_i)(\Delta_{ij}+\Delta_{ic}-U_i)}  \nonumber \\
    &\quad +\frac{2g_{ij}g_{ic}g_{jc}}{(\Delta_{ic}-U_i)} \left[\frac{1}{(\Delta_{ij}+\Delta_{ic}-U_i)}+\frac{1}{(\Delta_{ij}-U_i)} \right] \nonumber \\
    &\quad +\frac{4g_{ic}^2g_{jc}^2}{(\Delta_{ic}-U_i)(\Delta_{ij}+\Delta_{ic}-U_i)(2\Delta_{ic}+U_c-U_i)}\nonumber\\
    &\quad +\frac{2g_{ic}^2g_{jc}^2}{(\Delta_{ic}-U_i)(\Delta_{ij}+\Delta_{ic}-U_i)(\Delta_{ij}-U_i)} 
\end{align} 
Such that the effective hopping term is $J'_{20}=(\braket{2_1|\hat{H}|2_2}+\braket{2_2|\hat{H}|2_1}^{\star})/2 $. For the homogeneous configuration, consistent with Eq.~\eqref{J_200}, we get
\begin{align}
    J'_{20}
    =
    -\frac{2g_q^2}{U}
    -&\frac{2g_qg_c^2}{(\Delta_c-U)}\left[
    \frac{1}{(\Delta_c-U)}
    -\frac{2}{U} 
    \right] \nonumber \\
    &+\frac{2g_c^4}{(\Delta_c-U)^2}\left[
    \frac{1}{\Delta_c}
    -\frac{1}{U} 
    \right],
    \label{J_200_Perturbation}
\end{align}
which remains valid near $\Delta_{c}=-U$ although away from $U=0$. Both results \eqref{J_200} and \eqref{J_200_Perturbation} are similar away from resonance points, although \eqref{J_200_Perturbation} is more accurate due to the simplicity of the perturbative parameters involved. 

\subsection{$\ket{21} \leftrightarrow \ket{12}$}\label{App:couplers_21}
Second, we consider the case where there is a leakage in the first transmons and one excitation in the second transmon. We obtain the two-dimensional effective Hamiltonian given by
\begin{align}
    \frac{\hat{H}_{\text{eff}}}{\hbar}
    =
    \begin{pmatrix}
        (E_{21}^{(0)}+E_{21}^{(2)})/\hbar & 2 g_{12}+J^{(2)}+J^{(4)}\\
        2g_{12}+J^{(2)}+J^{(4)} & (E_{12}^{(0)}+E_{12}^{(2)})/\hbar
    \end{pmatrix},
\end{align}
where $E_{21}^{(0)}/\hbar =2\omega_1+\omega_2-U_1$ and $E_{12}^{(0)}/\hbar=2\omega_2+\omega_1-U_2 $. The second order corrections to the energy are
\begin{align}
    E_{21}^{(2)}=&\hbar g_{1c}^2\left( \frac{2}{\Delta_{1c}-U_1}+\frac{1}{\Delta_{2c}}\right), \\
    E_{12}^{(2)}=& \hbar g_{2c}^2\left( \frac{1}{\Delta_{1c}}+\frac{2}{\Delta_{2c}-U_2}\right),
\end{align}
and the second order corrections to the hopping term is
\begin{align}
    J^{(2)}=g_{1c}g_{2c} \left(\frac{1}{\Delta_{1c}-U_1} +\frac{1}{\Delta_{2c}-U_2}\right).
    \label{J_201_2}
\end{align}
The fourth order correction is given by
\begin{widetext}
\begin{align}
    J^{(4)}=&g_{12}\left[  g_{1c}^2 \left(\frac{2}{\Delta_{1c}(\Delta_{1c}-U_1)}-\frac{1}{\Delta_{1c}^2}-\frac{2}{(\Delta_{1c}-U_1)^2}\right)+g_{2c}^2 \left(\frac{2}{\Delta_{2c}(\Delta_{2c}-U_2)}-\frac{1}{\Delta_{2c}^2}-\frac{2}{(\Delta_{2c}-U_2)^2}\right)\right] \nonumber\\
    &+\frac{g_{1c}g_{2c}}{8}
    \left[
    -\frac{4g_{1c}^2}{(\Delta_{1c}-U_1)^3}-\frac{4g_{2c}^2}{(\Delta_{2c}-U_2)^3}
    -\frac{28g_{1c}^2}{(\Delta_{1c}-U_1)^2(\Delta_{2c}-U_2)}-\frac{28g_{2c}^2}{(\Delta_{1c}-U_1)(\Delta_{2c}-U_2)^2} \right. \nonumber\\
    &\quad \qquad \qquad +\frac{24}{(\Delta_{1c}-U_1)(\Delta_{2c}-U_2)}\left(\frac{g_{1c}^2}{\Delta_{1c}+U_c}+\frac{g_{2c}^2}{\Delta_{2c}+U_c}\right) 
    -\frac{8}{(\Delta_{1c}-U_1)(\Delta_{2c}-U_2)}\left(\frac{g_{1c}^2}{\Delta_{1c}}+\frac{g_{2c}^2}{\Delta_{2c}}\right) \nonumber\\
    & \quad \qquad \qquad -4\left(\frac{g_{1c}^2}{(\Delta_{1c}+U_c)^2}+\frac{g_{2c}^2}{(\Delta_{2c}+U_c)^2}\right)\left(\frac{1}{\Delta_{1c}-U_1}+\frac{1}{\Delta_{2c}-U_2}\right) \nonumber \\
    &\quad \qquad \qquad -\frac{2g_{1c}^2}{\Delta_{1c}^2(\Delta_{2c}-U_2)}-\frac{2g_{2c}^2}{\Delta_{2c}^2(\Delta_{1c}-U_1)}
    -\frac{6g_{1c}^2}{\Delta_{1c}^2(\Delta_{1c}-U_1)}-\frac{6g_{2c}^2}{\Delta_{2c}^2(\Delta_{2c}-U_2)} \nonumber \\
    &\quad \qquad \qquad +\frac{12g_{2c}^2}{(\Delta_{1c}+U_c-U_1)\Delta_{2c}(\Delta_{2c}-U_2)}+\frac{12g_{1c}^2}{\Delta_{1c}(\Delta_{2c}+U_c-U_2)(\Delta_{1c}-U_1)}
 \nonumber \\
    &\quad \qquad \qquad -\frac{4g_{1c}^2}{\Delta_{1c}(\Delta_{2c}+U_c-U_2)(\Delta_{1c}+U_c)}-\frac{4g_{2c}^2}{(\Delta_{1c}+U_c-U_1)\Delta_{2c}(\Delta_{2c}+U_c)} \nonumber \\
    &\quad \qquad \qquad -\frac{4g_{1c}^2}{(\Delta_{2c}+U_c-U_2)(\Delta_{1c}-U_1)(\Delta_{1c}+U_c)}-\frac{4g_{2c}^2}{(\Delta_{1c}+U_c-U_1)(\Delta_{2c}-U_2)(\Delta_{2c}+U_c)} \nonumber \\
    &\quad \qquad \qquad \left.+\frac{12g_{1c}^2}{\Delta_{1c}(\Delta_{1c}-U_1)(\Delta_{1c}+U_c)}+\frac{12g_{2c}^2}{\Delta_{2c}(\Delta_{2c}-U_2)(\Delta_{2c}+U_c)} \right].
    \label{J_201_4}
\end{align}
\end{widetext}
Note that we have not take into account the states $ \ket{30}, \ket{03}$, since they participate at fourth order correcting the energies and at sixth order correcting the hoppings. This implies that we can describe the system as a one excitation oscillating between the two transmons with an effective hopping strength given by $J_{21}=2 g_{12}+J^{(2)}+J^{(4)}$. Due to the presence of the terms $(\Delta_{jc}+U_c)^{-1} $ and $(\Delta_{jc}+U_c)^{-2} $ in \eqref{J_201_4}, this fourth order correction is less accurate than the second order correction \eqref{J_201_2} near $\Delta_{jc}=-U$. However, the fourth order correction is more accurate in the regime $\Delta_{jc}>U$, and it is the one used to obtain the blocking condition.

\section{Three transmons and two tunable couplers}
\label{App:array_couplers}
In this longer array, we follow the same perturbative process as in the previous case. Since here the system has two tunable couplers, the $\hat{S}$ operator satisfies $[\hat{S}, \hat{H}_0] = -\hat{V}_{c_1}-\hat{V}_{c_2}$. So it can be expressed as a sum of two terms $ \hat{S}=\hat{S}_{c_1}+\hat{S}_{c_2}$, each of them corresponding to a different tunable coupler as expressed in Eq.~\eqref{SW_operator}. We get similar result as in the previous case, the difference lies in the middle transmons that get additional contributions to the corrections.

\subsection{$\ket{201} \leftrightarrow \ket{102}$}
For the case where there is a leakage in the first transmon and one excitation in the third transmon, we obtain the three-dimensional effective Hamiltonian
\begin{align}
    &\frac{\hat{H}_{\text{eff}}}{\hbar}=\notag \\
    &\quad \begin{pmatrix}
        \frac{E_{201}^{(0)}}{\hbar}+\frac{E_{201}^{(2)}}{\hbar} & \sqrt{2} g_{12}+J^{(2)}_1 & J^{(4)} \\
        \sqrt{2} g_{12}+J^{(2)}_1 & \frac{E_{111}^{(0)}}{\hbar}+\frac{E_{111}^{(2)}}{\hbar} & \sqrt{2} g_{23}+J^{(2)}_2 \\
        J^{(4)} & \sqrt{2} g_{23}+J^{(2)}_2 & \frac{E_{102}^{(0)}}{\hbar}+\frac{E_{102}^{(2)}}{\hbar}
    \end{pmatrix},
\end{align}
where $E_{201}^{(0)}/\hbar=2\omega_1+\omega_3-U_1$, $E_{111}^{(0)}/\hbar=\omega_1+\omega_2+\omega_3$, and $E_{102}^{(0)}/\hbar=2\omega_2+\omega_1-U_3$. The second order corrections to the energies are
\begin{align}
    \frac{E_{201}^{(2)}}{\hbar}=&  \frac{2g_{1c_1}^2}{\Delta_{1c_1}-U_1}+\frac{g_{3c_2}^2}{\Delta_{3c_2}}, \\
    \frac{E_{111}^{(2)}}{\hbar}=&  \frac{g_{1c_1}^2}{\Delta_{1c_1}}+\frac{g_{2c_1}^2}{\Delta_{2c_1}}+\frac{g_{2c_2}^2}{\Delta_{2c_2}}+\frac{g_{3c_2}^2}{\Delta_{3c_2}}, \\
    \frac{E_{102}^{(2)}}{\hbar}=&\frac{g_{1c_1}^2}{\Delta_{1c_1}}+\frac{2g_{3c_2}^2}{\Delta_{3c_2}-U_3},
\end{align}
and the second order corrections to the hopping terms are
\begin{align}
    J^{(2)}_1=&\sqrt{2}\frac{g_{1c_1}g_{2c_1}}{2} \left(\frac{1}{\Delta_{1c_1}-U_1} +\frac{1}{\Delta_{2c_1}}\right), \\
    J^{(2)}_2=&\sqrt{2}\frac{g_{2c_2}g_{3c_2}}{2} \left(\frac{1}{\Delta_{2c_2}} +\frac{1}{\Delta_{3c_2}-U_3}\right).
\end{align}
For the particular case $ \omega_1=\omega_3$, $g_q\equiv g_{12}=g_{23}$, $g_c\equiv g_{1c_1}=g_{3c_2}=-g_{2c_1}=-g_{2c_2}$, and homogeneous anharmonicity, we have $ J_1^{(2)}=J_2^{(2)}$ and the fourth order correction is given by
\begin{align}
    J^{(4)}=&\frac{2g_q g_c^2}{\Delta_c(\Delta_c-\Delta_q-U)}  \label{J_4_20001} \\
    &-\frac{g_c^4}{2 \Delta_c(\Delta_c-\Delta_q-U)} \left[\frac{1}{\Delta_c} +\frac{3}{\Delta_c-\Delta_q-U}\right].
\end{align}
Then, when the state $E_{111} $ is detuned with respect to $E_{201} $ and $E_{102} $, we can describe the system as a single excitation oscillating between the first and third transmon, with an effective hopping given by
\begin{align}
    J_{201}=\frac{\hbar \left(\sqrt{2}g_q+J^{(2)}\right)^2}{E_{201}^{(0)}-E_{111}^{(0)}}+J^{(4)}.
    \label{Jeff_20001}
\end{align}
Without taking into account second order correction to the energies we can obtain a simple expression for $J_{201}=0$ given by
\begin{align}
    \Delta_q \approx \Delta_c-U-\frac{g_c^2}{g_q}.
    \label{Jeff_20001_zero}
\end{align} 
Note that we have excluded in the whole analysis states involving three excitations at the same site since they are relevant only at higher orders.

\begin{figure*}
    \centering
    \includegraphics[width=1\linewidth]{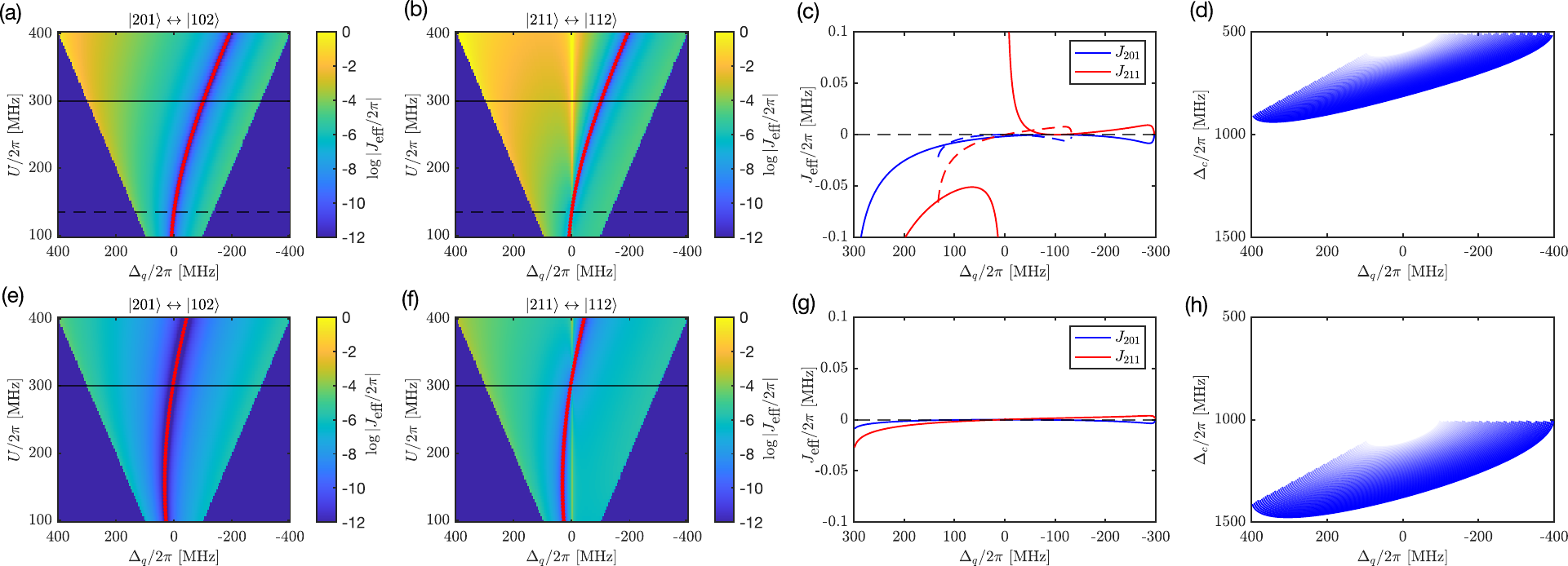}
    \caption{Effective hopping strengths $\ket{201} \leftrightarrow \ket{102}$ and $\ket{211} \leftrightarrow \ket{112}$ in the $ZZ$-OFF position as a function of the transmons detuning  $\Delta_q=\omega_2-\omega_1=\omega_2-\omega_3$ and anharmonicity $U$ for $g_c/2\pi=50$ MHz and $g_q/2\pi=5$ MHz (a-d), and $g_c/2\pi \approx 63.25$ MHz and $g_q/2\pi=4$ MHz (e-h). (a,e) Effective hopping strengths for the case \eqref{Jeff_20001}. (b,f) Effective hopping strengths for the case \eqref{Jeff_20101}. Black solid and dashed lines indicate $U/2\pi=300$ MHz and $U/2\pi=135$ MHz, respectively. Since in each case the $ZZ$-OFF condition is satisfied, the range of the detuning is limited $\Delta_q=[-U,+U]$. (c,g) Effective hopping strengths for $U/2\pi=300$ MHz (solid lines) and $U/2\pi=135$ MHz (dashed lines). The red solid lines correspond to the zero condition of Eq. \eqref{Jeff_20001_zero}. (d,h) Transmon-couplers detuning $\Delta_c=\omega_2-\omega_c$ for each transmons detuning $\Delta_q$ fulfilling the $ZZ$-off condition. Each line represent a value of the anharmonicity $U/2\pi$, from $U/2\pi=100$ MHz (light blue) to $U/2\pi=400$ MHz (dark blue). The frequency of the middle transmon is $\omega_2/2\pi=4000$ MHz.  }
    \label{fig:map_tunneling_zeros}
\end{figure*}

\subsection{$\ket{211} \leftrightarrow \ket{112}$}
For the case where there is a leakage in the first transmon and one excitation in the second and third transmon, we obtain the four-dimensional effective Hamiltonian
\begin{align}
    &\frac{\hat{H}_{\text{eff}}}{\hbar}=\\
    &\begin{pmatrix}
        \frac{E_{211}^{(0)}}{\hbar}+\frac{E_{211}^{(2)}}{\hbar} & 2g_{12}+J^{(2)}_1 & \sqrt{2}g_{23}+J_2^{(2)} & J^{(4)}\\
        2g_{12}+J^{(2)}_1 & \frac{E_{121}^{(0)}}{\hbar}+\frac{E_{121}^{(2)}}{\hbar} & \mathcal{O}(\lambda^3 g_c) & 2g_{23}+J_3^{(2)}\\
        \sqrt{2}g_{23}+J_2^{(2)} & \mathcal{O}(\lambda^3 g_c) & \frac{E_{202}^{(0)}}{\hbar}+\frac{E_{202}^{(2)}}{\hbar} & \sqrt{2}g_{12}+J^{(2)}_4\\
        J^{(4)} & 2g_{23}+J_3^{(2)} & \sqrt{2}g_{12}+J^{(2)}_4 & \frac{E_{112}^{(0)}}{\hbar}+\frac{E_{112}^{(2)}}{\hbar}
    \end{pmatrix},\notag
\end{align}
where $E_{211}^{(0)}/\hbar=2\omega_1+\omega_2+\omega_3-U_1$, $E_{121}^{(0)}/\hbar=\omega_1+2\omega_2+\omega_3-U_2$, $E_{202}^{(0)}/\hbar=2\omega_1+2\omega_3-U_1-U_3 $, and $E_{112}^{(0)}/\hbar=\omega_1+\omega_2+2\omega_3-U_3 $. We have denoted as $\mathcal{O}(\lambda^3 g_c)$ the terms involving transition between intermediate states, which we are not going to consider here. The second order corrections to the energies are
\begin{align}
    \frac{E_{211}^{(2)}}{\hbar}=& \frac{2g_{1c_1}^2}{\Delta_{1c_1}-U_1}+\frac{g_{2c_1}^2}{\Delta_{2c_1}}+\frac{g_{2c_2}^2}{\Delta_{2c_2}}+\frac{g_{3c_2}^2}{\Delta_{3c_2}}, \\
    \frac{E_{121}^{(2)}}{\hbar}=& \frac{g_{1c_1}^2}{\Delta_{1c_1}}+\frac{2g_{2c_1}^2}{\Delta_{2c_1}-U_2}+\frac{2g_{2c_2}^2}{\Delta_{2c_2}-U_2}+\frac{g_{3c_2}^2}{\Delta_{3c_2}}, \\
    \frac{E_{202}^{(2)}}{\hbar}=& \frac{2g_{1c_1}^2}{\Delta_{1c_1}-U_1}+\frac{2g_{3c_2}^2}{\Delta_{3c_2}-U_3}, \\
    \frac{E_{112}^{(2)}}{\hbar}=&\frac{g_{1c_1}^2}{\Delta_{1c_1}}+\frac{g_{2c_1}^2}{\Delta_{2c_1}}+\frac{g_{2c_2}^2}{\Delta_{2c_2}}+\frac{2g_{3c_2}^2}{\Delta_{3c_2}-U_3},
\end{align}
and the second order corrections to the hopping terms are
\begin{align}
    J^{(2)}_1=&g_{1c_1}g_{2c_1} \left(\frac{1}{\Delta_{1c_1}-U_1} +\frac{1}{\Delta_{2c_1}-U_2}\right), \\
    J^{(2)}_2=&\frac{g_{2c_2}g_{3c_2}}{\sqrt{2}} \left(\frac{1}{\Delta_{2c_2}} +\frac{1}{\Delta_{3c_2}-U_3}\right), \\
    J^{(2)}_3=&g_{2c_2}g_{3c_2} \left(\frac{1}{\Delta_{2c_2}-U_2} +\frac{1}{\Delta_{3c_2}-U_3}\right), \\
    J^{(2)}_4=&\frac{g_{1c_1}g_{2c_1}}{\sqrt{2}} \left(\frac{1}{\Delta_{1c_1}-U_1}+\frac{1}{\Delta_{2c_1}} \right).
\end{align}
As in the previous case, for $ \omega_1=\omega_3$, $g_q\equiv g_{12}=g_{23}$, $g_c\equiv g_{1c_1}=g_{3c_2}=-g_{2c_1}=-g_{2c_2}$, and homogeneous anharmonicity, we have $ J_1^{(2)}=J_3^{(2)}$ and $ J_2^{(2)}=J_4^{(2)}$ and the fourth order correction is given by
\begin{align}
    J^{(4)}=&\frac{g_q g_c^2}{\Delta_c-\Delta_q-U}\left[ \frac{4}{\Delta_c-U}-\frac{2}{\Delta_c}\right]     \label{J_4_20101} \\
    &+\frac{g_c^4}{2 (\Delta_c-\Delta_q-U)} \left[\frac{1}{\Delta_c^2} +\frac{3}{\Delta_c(\Delta_c-\Delta_q-U)} \right. \nonumber \\
    &\left.-\frac{2}{(\Delta_c-U)^2}-\frac{6}{(\Delta_c-U)(\Delta_c-\Delta_q-U)}\right]. \notag 
\end{align}
In the same sense as before, we can describe the dynamics as the oscillations of one excitation between the first and third transmon, when the energy of $E_{211} $ and $E_{112} $ are different from $E_{121} $ and $E_{202} $. Then, we can obtain an effective hopping 
\begin{align}
    J_{211}=\frac{\hbar \left(2g_q+J_1^{(2)}\right)^2}{E_{211}^{(0)}-E_{121}^{(0)}}+\frac{\hbar \left(\sqrt{2}g_q+J_2^{(2)}\right)^2}{E_{211}^{(0)}-E_{202}^{(0)}}+J^{(4)}.
    \label{Jeff_20101}
\end{align}
In this case, the expression for the effective hopping strength becomes significantly more intricate. Nevertheless, provided the relative scales of the physical parameters are preserved to maintain the qualitative features of the effective hopping landscape shown in Fig.~\ref{fig:map_tunneling_zeros}, we can exploit the structural similarities between Eq.~\eqref{Jeff_20001} and Eq.~\eqref{Jeff_20101}. Away from the central resonance at $\Delta_q = 0$, both expressions are functionally identical, with the sole exception of the terms depending on $(\Delta_c - U)^{-1}$ and $(\Delta_c - U)^{-2}$ that appear explicitly in Eq.~\eqref{J_4_20101}. Consequently, the zero-cancellation condition established in Eq.~\eqref{Jeff_20001_zero} serves as a reliable reference. This analytical approximation breaks down near $\Delta_q = 0$ and exhibits reduced accuracy in the $\Delta_q < 0$ region as the system approaches the $\Delta_c = U$ resonance. Note that close to the resonances, the system is inherently three-dimensional.

As demonstrated, the analytical solution~\eqref{Jeff_20001_zero} explicitly depends on the transmon anharmonicity $U$ and the coupling parameter ratio $g_c^2/g_q$. In Fig.~\ref{fig:map_tunneling_zeros}, we evaluate the effective hopping strengths and the zero-coupling conditions across various values of anharmonicity for two distinct parameter sets: $g_c/2\pi = 50$~MHz and $g_q = 5$~MHz [Figs.~\ref{fig:map_tunneling_zeros}(a-d)], and $g_c/2\pi =20\sqrt{10}$~MHz and $g_q = 4$~MHz [Figs.~\ref{fig:map_tunneling_zeros}(e-h)]. Crucially, these coupling parameters can only be varied within a narrow window to preserve both the validity of the perturbative expansion and the dynamics of interest derived above. Furthermore, because we evaluate the tunneling dynamics specifically at the $ZZ$-OFF operating point, we must map the coupler frequency profile as a function of the transmon detuning, analogous to Fig.~\ref{fig:effective_coupling}(a), for each unique parameter set. In these frequency plots, each color gradient-coded line represents a specific anharmonicity value corresponding to a fixed ratio of $(g_c^2/g_q) = 2\pi \times 500$~MHz [Fig.~\ref{fig:map_tunneling_zeros}(d)] or $(g_c^2/g_q) =2\pi\times 1000$~MHz [Fig.~\ref{fig:map_tunneling_zeros}(h)].

Figures ~\ref{fig:map_tunneling_zeros}(a, b) illustrate the results for the parameters established in the main text $g_c/2\pi = 50$~MHz and $g_q = 5$~MHz, where the solid black lines correspond to our standard anharmonicity value of $U/2\pi = 300$~MHz. Under this configuration, if leakage populates transmon A, the zero-coupling blocking condition for an ABA array is met at the operating frequencies $\omega_{\text{A}}/2\pi = 4101$~MHz and $\omega_{\text{B}}/2\pi = 4000$~MHz. However, in an extended ABAB array, a leakage excitation localized at a B site can still tunnel to adjacent B transmons. The resulting hopping strengths scale to $|J_{201}|/2\pi \approx 8$~kHz and $|J_{211}|/2\pi \approx 60$~kHz, as shown by the colored solid lines in Fig.~\ref{fig:map_tunneling_zeros}(c). 

If the anharmonicity is instead reduced to $U/2\pi = 135$~MHz [represented by the dashed black lines in Figs.~\ref{fig:map_tunneling_zeros}(a, b)], the zero-coupling condition shifts exactly to zero detuning $\Delta_q = 0$, which implies a uniform BBB array configuration. Because this symmetric configuration can introduce unwanted phenomena, we can deliberately choose a safe operational detuning, such as $\omega_{\text{A}}/2\pi = 4040$~MHz and $\omega_{\text{B}}/2\pi = 4000$~MHz. At this biased operating point, the system is moved away from the exact zero-interaction condition but successfully achieves highly suppressed, homogeneous tunneling strengths across the array, given by $|J_{201}|/2\pi \approx 0.7$~kHz and $|J_{211}|/2\pi \approx 5$~kHz (colored solid lines in Fig.~\ref{fig:map_tunneling_zeros}(c)).

This interaction landscape can be optimized further by tuning the bare coupling strengths. In Figs.~\ref{fig:map_tunneling_zeros}(e, f), we slightly modify the parameters to $g_c/2\pi  = 20 \sqrt{10}$~MHz and $g_q/2\pi = 4$~MHz, while maintaining the baseline anharmonicity at $U/2\pi = 300$~MHz (solid black line). Here, the zero-interaction condition is again located at zero detuning. However, by introducing a small operational detuning of $\Delta_q/2\pi = \pm 40$~MHz, we obtain low and homogeneous tunneling strengths of $|J_{201}|/2\pi \approx 0.05$~kHz and $|J_{211}|/2\pi \approx 0.7$~kHz [Fig.~\ref{fig:map_tunneling_zeros}(g)].

\newpage 

\bibliography{references}

@article{wesdorp26,
      title={Mitigating crosstalk errors for simultaneous single-qubit gates on a superconducting quantum processor}, 
      author={Jaap J. Wesdorp and Eric Hyyppä and Joona Andersson and Janos Adam and Rohit Beriwal and Ville Bergholm and Saga Dahl and Simone Diego Fasciati and Alejandro Gomez Friero and Zheming Gao and others},
      year={2026},
      journal={arXiv:2603.11018},
      volume = {},
      pages = {},
      url={https://arxiv.org/abs/2603.11018}, 
}

@article{Karamlou24,
	title = {Probing entanglement in a {2D} hard-core {Bose}–{Hubbard} lattice},
	issn = {1476-4687},
	doi = {10.1038/s41586-024-07325-z},
	journal = {Nature},
	author = {Karamlou, Amir H. and Rosen, Ilan T. and Muschinske, Sarah E. and Barrett, Cora N. and Di Paolo, Agustin and Ding, Leon and Harrington, Patrick M. and Hays, Max and Das, Rabindra and Kim, David K. and Niedzielski, Bethany M. and Schuldt, Meghan and Serniak, Kyle and Schwartz, Mollie E. and Yoder, Jonilyn L. and Gustavsson, Simon and Yanay, Yariv and Grover, Jeffrey A. and Oliver, William D.},
	month = apr,
	year = {2024},
        volume = {629},
	pages = {561},
}

@article{Kjaergaard20,
	title = {Superconducting {Qubits}: {Current} {State} of {Play}},
	volume = {11},
	issn = {1947-5454},
	doi = {10.1146/annurev-conmatphys-031119-050605},
	number = {1},
	urldate = {2020-05-13},
	journal = {Ann. Rev. Cond. Matt. Phys.},
	publisher = {Annual Reviews},
	author = {Kjaergaard, Morten and Schwartz, Mollie E. and Braumüller, Jochen and Krantz, Philip and Wang, Joel I.-J. and Gustavsson, Simon and Oliver, William D.},
	month = mar,
	year = {2020},
	pages = {369--395},
}

@article{Rosenfeld24,
	title = {High-{Fidelity} {Two}-{Qubit} {Gates} between {Fluxonium} {Qubits} with a {Resonator} {Coupler}},
	volume = {5},
	doi = {10.1103/PRXQuantum.5.040317},
	number = {4},
	urldate = {2024-12-05},
	journal = {PRX Quantum},
	author = {Rosenfeld, Emma L. and Hann, Connor T. and Schuster, David I. and Matheny, Matthew H. and Clerk, Aashish A.},
	month = nov,
	year = {2024},
	pages = {040317},
}

@article{Zhang24,
  title = {Tunable Inductive Coupler for High-Fidelity Gates Between Fluxonium Qubits},
  author = {Zhang, Helin and Ding, Chunyang and Weiss, D.K. and Huang, Ziwen and Ma, Yuwei and Guinn, Charles and Sussman, Sara and Chitta, Sai Pavan and Chen, Danyang and Houck, Andrew A. and Koch, Jens and Schuster, David I.},
  journal = {PRX Quantum},
  volume = {5},
  issue = {2},
  pages = {020326},
  numpages = {18},
  year = {2024},
  month = {May},
  publisher = {American Physical Society},
  doi = {10.1103/PRXQuantum.5.020326},
  url = {https://link.aps.org/doi/10.1103/PRXQuantum.5.020326}
}

@article{Ding23,
	title = {High-{Fidelity}, {Frequency}-{Flexible} {Two}-{Qubit} {Fluxonium} {Gates} with a {Transmon} {Coupler}},
	volume = {13},
	url = {https://link.aps.org/doi/10.1103/PhysRevX.13.031035},
	doi = {10.1103/PhysRevX.13.031035},
	number = {3},
	urldate = {2024-04-29},
	journal = {Phys. Rev. X},
	publisher = {American Physical Society},
	author = {Ding, Leon and Hays, Max and Sung, Youngkyu and Kannan, Bharath and An, Junyoung and Di Paolo, Agustin and Karamlou, Amir H. and Hazard, Thomas M. and Azar, Kate and Kim, David K. and others},
	month = sep,
	year = {2023},
	pages = {031035},
}

@article{Motzoi09,
  title = {Simple Pulses for Elimination of Leakage in Weakly Nonlinear Qubits},
  author = {Motzoi, F. and Gambetta, J. M. and Rebentrost, P. and Wilhelm, F. K.},
  journal = {Phys. Rev. Lett.},
  volume = {103},
  issue = {11},
  pages = {110501},
  numpages = {4},
  year = {2009},
  month = {Sep},
  publisher = {American Physical Society},
  doi = {10.1103/PhysRevLett.103.110501},
  url = {https://link.aps.org/doi/10.1103/PhysRevLett.103.110501}
}

@article{Hyyppa24,
  title = {Reducing Leakage of Single-Qubit Gates for Superconducting Quantum Processors Using Analytical Control Pulse Envelopes},
  author = {Hyyppä, Eric and Vepsäläinen, Antti and Papič, Miha and Chan, Chun Fai and Inel, Sinan and Landra, Alessandro and Liu, Wei and Luus, Jürgen and Marxer, Fabian and Ockeloen-Korppi, Caspar and Orbell, Sebastian and Tarasinski, Brian and Heinsoo, Johannes},
  journal = {PRX Quantum},
  volume = {5},
  issue = {3},
  pages = {030353},
  numpages = {28},
  year = {2024},
  month = {Sep},
  publisher = {American Physical Society},
  doi = {10.1103/PRXQuantum.5.030353},
  url = {https://link.aps.org/doi/10.1103/PRXQuantum.5.030353}
}

@article{Marques23,
  title = {All-Microwave Leakage Reduction Units for Quantum Error Correction with Superconducting Transmon Qubits},
    author = {Marques, J. F. and Ali, H. and Varbanov, B. M. and Finkel, M. and Veen, H. M. and van der Meer, S. L. M. and Valles-Sanclemente, S. and Muthusubramanian, N. and Beekman, M. and Haider, N. and Terhal, B. M. and DiCarlo, L.},
  journal = {Phys. Rev. Lett.},
  volume = {130},
  issue = {25},
  pages = {250602},
  numpages = {6},
  year = {2023},
  month = {Jun},
  publisher = {American Physical Society},
  doi = {10.1103/PhysRevLett.130.250602},
  url = {https://link.aps.org/doi/10.1103/PhysRevLett.130.250602}
}

@Article{varbanov20,
author={Varbanov, Boris Mihailov
and Battistel, Francesco
and Tarasinski, Brian Michael
and Ostroukh, Viacheslav Petrovych
and O'Brien, Thomas Eugene
and DiCarlo, Leonardo
and Terhal, Barbara Maria},
title={Leakage detection for a transmon-based surface code},
journal={npj Quantum Inf.},
year={2020},
month={Dec},
day={14},
volume={6},
number={1},
pages={102},
issn={2056-6387},
doi={10.1038/s41534-020-00330-w},
url={https://doi.org/10.1038/s41534-020-00330-w}
}

@article{Marxer26,
	title = {Above 99.9\% {Fidelity} {Single}-{Qubit} {Gates}, {Two}-{Qubit} {Gates}, and {Readout} in a {Single} {Superconducting} {Quantum} {Device}},
	volume = {7},
	issn = {2691-3399},
	doi = {10.1103/n86s-2b88},
	number = {2},
	urldate = {2026-06-13},
	journal = {PRX Quantum},
	author = {Marxer, Fabian and Mrożek, Jakub and Andersson, Joona and Abdurakhimov, Leonid and Adam, Janos and Bergholm, Ville and Beriwal, Rohit and Chan, Chun Fai and Dahl, Saga and Das, Soumya Ranjan and others}, 
	month = may,
	year = {2026},
	pages = {020333},
}

@article{Terhal15,
	title = {Quantum error correction for quantum memories},
	volume = {87},
	url = {http://link.aps.org/doi/10.1103/RevModPhys.87.307},
	doi = {10.1103/RevModPhys.87.307},
	number = {2},
	urldate = {2015-04-28},
	journal = {Rev. Mod. Phys.},
	author = {Terhal, Barbara M.},
	month = apr,
	year = {2015},
	pages = {307--346},
}

@article{Fowler13,
  title = {Coping with qubit leakage in topological codes},
  author = {Fowler, Austin G.},
  journal = {Phys. Rev. A},
  volume = {88},
  issue = {4},
  pages = {042308},
  numpages = {5},
  year = {2013},
  month = {Oct},
  publisher = {American Physical Society},
  doi = {10.1103/PhysRevA.88.042308},
  url = {https://link.aps.org/doi/10.1103/PhysRevA.88.042308}
}

@article{Camps24,
	title = {Leakage {Mobility} in {Superconducting} {Qubits} as a {Leakage} {Reduction} {Unit}},
	doi = {10.1088/1402-4896/ae4dc0},
	journal = {Phys. Scr.},
	author = {Camps, Joan and Crawford, Ophelia and Gehér, György P. and Gramolin, Alexander V. and Stafford, Matthew P. and Turner, Mark},
	month = jun,
	year = {2026},
        volume = {101},
        pages = {115107},
}

@article{martinvazquez_2025,
  title = {Passive leakage removal unit based on a disordered transmon array},
  author = {Mart\'{\i}n-V\'azquez, Gonzalo and Tolppanen, Taneli and Silveri, Matti},
  journal = {Phys. Rev. A},
  volume = {112},
  issue = {3},
  pages = {032408},
  numpages = {23},
  year = {2025},
  month = {Sep},
  publisher = {American Physical Society},
  doi = {10.1103/bxqs-y662},
  url = {https://link.aps.org/doi/10.1103/bxqs-y662}
}

@article{sung_2021,
  title = {Realization of High-Fidelity {CZ} and {ZZ}-Free {iSWAP} Gates with a Tunable Coupler},
  author = {Sung, Youngkyu and Ding, Leon and Braum\"uller, Jochen and Veps\"al\"ainen, Antti and Kannan, Bharath and Kjaergaard, Morten and Greene, Ami and Samach, Gabriel O. and McNally, Chris and Kim, David and others},
  journal = {Phys. Rev. X},
  volume = {11},
  issue = {2},
  pages = {021058},
  numpages = {32},
  year = {2021},
  month = {Jun},
  publisher = {American Physical Society},
  doi = {10.1103/PhysRevX.11.021058},
  url = {https://link.aps.org/doi/10.1103/PhysRevX.11.021058}
}

@article{chu_2021,
  title = {Coupler-Assisted Controlled-Phase Gate with Enhanced Adiabaticity},
  author = {Chu, Ji and Yan, Fei},
  journal = {Phys. Rev. Appl.},
  volume = {16},
  issue = {5},
  pages = {054020},
  numpages = {21},
  year = {2021},
  month = {Nov},
  publisher = {American Physical Society},
  doi = {10.1103/PhysRevApplied.16.054020},
  url = {https://link.aps.org/doi/10.1103/PhysRevApplied.16.054020}
}

@article{yan_2018,
  title = {Tunable Coupling Scheme for Implementing High-Fidelity Two-Qubit Gates},
  author = {Yan, Fei and Krantz, Philip and Sung, Youngkyu and Kjaergaard, Morten and Campbell, Daniel L. and Orlando, Terry P. and Gustavsson, Simon and Oliver, William D.},
  journal = {Phys. Rev. Appl.},
  volume = {10},
  issue = {5},
  pages = {054062},
  numpages = {9},
  year = {2018},
  month = {Nov},
  publisher = {American Physical Society},
  doi = {10.1103/PhysRevApplied.10.054062},
  url = {https://link.aps.org/doi/10.1103/PhysRevApplied.10.054062}
}

@article{marxer_2023,
  title = {Long-Distance Transmon Coupler with {CZ}-Gate Fidelity above {99.8\%}},
  author = {Marxer, Fabian and Veps\"al\"ainen, Antti and Jolin, Shan W. and Tuorila, Jani and Landra, Alessandro and Ockeloen-Korppi, Caspar and Liu, Wei and Ahonen, Olli and Auer, Adrian and others}, 
  journal = {PRX Quantum},
  volume = {4},
  issue = {1},
  pages = {010314},
  numpages = {23},
  year = {2023},
  month = {Feb},
  publisher = {American Physical Society},
  doi = {10.1103/PRXQuantum.4.010314},
  url = {https://link.aps.org/doi/10.1103/PRXQuantum.4.010314}
}

@article{chen_2014,
  title = {Qubit Architecture with High Coherence and Fast Tunable Coupling},
  author = {Chen, Yu and Neill, C. and Roushan, P. and Leung, N. and Fang, M. and Barends, R. and Kelly, J. and Campbell, B. and Chen, Z. and Chiaro, B. and others},
  journal = {Phys. Rev. Lett.},
  volume = {113},
  issue = {22},
  pages = {220502},
  numpages = {5},
  year = {2014},
  month = {Nov},
  publisher = {American Physical Society},
  doi = {10.1103/PhysRevLett.113.220502},
  url = {https://link.aps.org/doi/10.1103/PhysRevLett.113.220502}
}

@article{eyob_2021,
  title = {Floating Tunable Coupler for Scalable Quantum Computing Architectures},
  author = {Sete, Eyob A. and Chen, Angela Q. and Manenti, Riccardo and Kulshreshtha, Shobhan and Poletto, Stefano},
  journal = {Phys. Rev. Appl.},
  volume = {15},
  issue = {6},
  pages = {064063},
  numpages = {12},
  year = {2021},
  month = {Jun},
  publisher = {American Physical Society},
  doi = {10.1103/PhysRevApplied.15.064063},
  url = {https://link.aps.org/doi/10.1103/PhysRevApplied.15.064063}
}

@article{li_2020,
  title = {Tunable Coupler for Realizing a Controlled-Phase Gate with Dynamically Decoupled Regime in a Superconducting Circuit},
  author = {Li, X. and Cai, T. and Yan, H. and Wang, Z. and Pan, X. and Ma, Y. and Cai, W. and Han, J. and Hua, Z. and Han, X. and Wu, Y. and Zhang, H. and Wang, H. and Song, Yipu and Duan, Luming and Sun, Luyan},
  journal = {Phys. Rev. Appl.},
  volume = {14},
  issue = {2},
  pages = {024070},
  numpages = {14},
  year = {2020},
  month = {Aug},
  publisher = {American Physical Society},
  doi = {10.1103/PhysRevApplied.14.024070},
  url = {https://link.aps.org/doi/10.1103/PhysRevApplied.14.024070}
}

@article{bravyi_2011,
title = {Schrieffer–{Wolff} transformation for quantum many-body systems},
journal = {Ann. Phys.},
volume = {326},
number = {10},
pages = {2793-2826},
year = {2011},
issn = {0003-4916},
doi = {https://doi.org/10.1016/j.aop.2011.06.004},
url = {https://www.sciencedirect.com/science/article/pii/S0003491611001059},
author = {Sergey Bravyi and David P. DiVincenzo and Daniel Loss}
}

@article{stehlik_2021,
  title = {Tunable Coupling Architecture for Fixed-Frequency Transmon Superconducting Qubits},
  author = {Stehlik, J. and Zajac, D. M. and Underwood, D. L. and Phung, T. and Blair, J. and Carnevale, S. and Klaus, D. and Keefe, G. A. and Carniol, A. and Kumph, M. and Steffen, Matthias and Dial, O. E.},
  journal = {Phys. Rev. Lett.},
  volume = {127},
  issue = {8},
  pages = {080505},
  numpages = {6},
  year = {2021},
  month = {Aug},
  publisher = {American Physical Society},
  doi = {10.1103/PhysRevLett.127.080505},
  url = {https://link.aps.org/doi/10.1103/PhysRevLett.127.080505}
}

@article{espinos_2023,
doi = {10.1088/2058-9565/acbed7},
url = {https://doi.org/10.1088/2058-9565/acbed7},
year = {2023},
month = {mar},
publisher = {IOP Publishing},
volume = {8},
number = {2},
pages = {025017},
author = {Espinós, H and Panadero, I and García-Ripoll, J J and Torrontegui, E},
title = {Quantum control of tunable-coupling transmons using dynamical invariants of motion},
journal = {Quantum Sci. Technol.}
}

@article{magesan_2020,
  title = {Effective Hamiltonian models of the cross-resonance gate},
  author = {Magesan, Easwar and Gambetta, Jay M.},
  journal = {Phys. Rev. A},
  volume = {101},
  issue = {5},
  pages = {052308},
  numpages = {15},
  year = {2020},
  month = {May},
  publisher = {American Physical Society},
  doi = {10.1103/PhysRevA.101.052308},
  url = {https://link.aps.org/doi/10.1103/PhysRevA.101.052308}
}

@article{kandala_2021,
  title = {Demonstration of a High-Fidelity {CNOT} Gate for Fixed-Frequency Transmons with Engineered {ZZ} Suppression},
  author = {Kandala, A. and Wei, K. X. and Srinivasan, S. and Magesan, E. and Carnevale, S. and Keefe, G. A. and Klaus, D. and Dial, O. and McKay, D. C.},
  journal = {Phys. Rev. Lett.},
  volume = {127},
  issue = {13},
  pages = {130501},
  numpages = {6},
  year = {2021},
  month = {Sep},
  publisher = {American Physical Society},
  doi = {10.1103/PhysRevLett.127.130501},
  url = {https://link.aps.org/doi/10.1103/PhysRevLett.127.130501}
}

@article{sanclemente_2026,
  title = {Optimizing the Frequency Positioning of Tunable Couplers in a Circuit {QED} Processor to Mitigate Spectator Effects on Quantum Operations},
  author = {Vall\'es-Sanclemente, S. and Vroomans, T. H. F. and van Abswoude, T. R. and Stavenga, T. and Brulleman, F. and van der Meer, S. L. M. and Xin, Y. and Lawrence, A. and Singh, V. and Rol, M. A. and DiCarlo, L.},
  journal = {Phys. Rev. Lett.},
  volume = {136},
  issue = {20},
  pages = {200801},
  numpages = {7},
  year = {2026},
  month = {May},
  publisher = {American Physical Society},
  doi = {10.1103/wdfj-4nbt},
  url = {https://link.aps.org/doi/10.1103/wdfj-4nbt}
}

@article{miao_overcoming_2023,
	title = {Overcoming leakage in quantum error correction},
	volume = {19},
	copyright = {2023 The Author(s)},
	issn = {1745-2481},
	url = {https://www.nature.com/articles/s41567-023-02226-w},
	doi = {10.1038/s41567-023-02226-w},
	number = {12},
	urldate = {2025-05-09},
	journal = {Nat. Phys.},
	publisher = {Nature Publishing Group},
	author = {Miao, Kevin C. and McEwen, Matt and Atalaya, Juan and Kafri, Dvir and Pryadko, Leonid P. and Bengtsson, Andreas and Opremcak, Alex and Satzinger, Kevin J. and Chen, Zijun and Klimov, Paul V. and others},
	month = dec,
	year = {2023},
	pages = {1780--1786},
}

@article{chapple_balanced_2025,
	title = {Balanced Cross-{Kerr} Coupling for Superconducting Qubit Readout},
	author = {Chapple, Alex A. and Benhayoune-Khadraoui, Othmane and Richer, Simon and Blais, Alexandre},
	journal = {Phys. Rev. Lett.},
	volume = {135},
	number = {25},
	pages = {256002},
	year = {2025},
	doi = {10.1103/r4v5-wyyt},
}

@article{beaulieu_fast_2026,
	title = {Fast, high-fidelity Transmon readout with intrinsic Purcell protection via nonperturbative cross-Kerr coupling},
	author = {Beaulieu, Guillaume and Chen, Jun-Zhe and Scigliuzzo, Marco and Benhayoune-Khadraoui, Othmane and Chapple, Alex A. and Spring, Peter A. and Blais, Alexandre and Scarlino, Pasquale},
	year = {2026},
	journal= {arXiv:2601.04975},
	pages = {},
        volume = {},
        doi = {10.48550/arXiv.2601.04975},
}

@article{wang_longitudinal_2025,
	title = {Longitudinal and Nonlinear Coupling for High-Fidelity Readout of a Superconducting Qubit},
	author = {Wang, Can and Liu, Feng-Ming and Chen, He and Du, Yi-Fei and Ying, Chong and Wang, Jian-Wen and Huo, Yong-Heng and Peng, Cheng-Zhi and Zhu, Xiaobo and Chen, Ming-Cheng and Lu, Chao-Yang and Pan, Jian-Wei},
	journal = {Phys. Rev. Lett.},
	volume = {135},
	number = {6},
	pages = {060803},
	year = {2025},
	doi = {10.1103/98n9-13y4},
}

@article{dumas_measurement-induced_2024,
	title = {Measurement-Induced Transmon Ionization},
	author = {Dumas, Marie Fr{\'e}d{\'e}rique and Groleau-Par{\'e}, Benjamin and {McDonald}, Alexander and Mu{\~n}oz-Arias, Manuel H. and Lled{\'o}, Crist{\'o}bal and D'Anjou, Benjamin and Blais, Alexandre},
	journal = {Phys. Rev. X},
	volume = {14},
	number = {4},
	pages = {041023},
	year = {2024},
	doi = {10.1103/PhysRevX.14.041023},
}

@article{khezri_measurement-induced_2023,
	title = {Measurement-induced state transitions in a superconducting qubit: Within the rotating-wave approximation},
	author = {Khezri, Mostafa and Opremcak, Alex and Chen, Zijun and Miao, Kevin C. and {McEwen}, Matt and Bengtsson, Andreas and White, Theodore and Naaman, Ofer and Sank, Daniel and Korotkov, Alexander N. and Chen, Yu and Smelyanskiy, Vadim},
	journal = {Phys. Rev. Appl.},
	volume = {20},
	number = {5},
	pages = {054008},
	year = {2023},
	doi = {10.1103/PhysRevApplied.20.054008},
}

@article{cohen_reminiscence_2023,
	title = {Reminiscence of Classical Chaos in Driven Transmons},
	author = {Cohen, Joachim and Petrescu, Alexandru and Shillito, Ross and Blais, Alexandre},
	journal = {PRX Quantum},
	volume = {4},
	number = {2},
	pages = {020312},
	year = {2023},
	doi = {10.1103/PRXQuantum.4.020312},
}

@article{sank_measurement-induced_2016,
	title = {Measurement-Induced State Transitions in a Superconducting Qubit: Beyond the Rotating Wave Approximation},
	author = {Sank, Daniel and Chen, Zijun and Khezri, Mostafa and Kelly, J. and Barends, R. and Campbell, B. and Chen, Y. and Chiaro, B. and Dunsworth, A. and Fowler, A. and Jeffrey, E. and Lucero, E. and Megrant, A. and Mutus, J. and Neeley, M. and others},
	journal = {Phys. Rev. Lett.},
	volume = {117},
	number = {19},
	pages = {190503},
	year = {2016},
	doi = {10.1103/PhysRevLett.117.190503},
}

@article{shillito_dynamics_2022,
	title = {Dynamics of Transmon Ionization},
	author = {Shillito, Ross and Petrescu, Alexandru and Cohen, Joachim and Beall, Jackson and Hauru, Markus and Ganahl, Martin and Lewis, Adam G.M. and Vidal, Guifre and Blais, Alexandre},
	journal = {Phys. Rev. Appl.},
	volume = {18},
	number = {3},
	pages = {034031},
	year = {2022},
	doi = {10.1103/PhysRevApplied.18.034031},
}

@article{blais_circuit_2021,
	title = {Circuit quantum electrodynamics},
	author = {Blais, Alexandre and Grimsmo, Arne L. and Girvin, S.M. and Wallraff, Andreas},
	journal = {Rev. Mod. Phys.},
	volume = {93},
	number = {2},
	pages = {025005},
	year = {2021},
	doi = {10.1103/RevModPhys.93.025005},
}

@article{martinvazquez_2024,
  title = {Phase transitions induced by standard and feedback measurements in transmon arrays},
  author = {Mart\'{\i}n-V\'azquez, Gonzalo and Tolppanen, Taneli and Silveri, Matti},
  journal = {Phys. Rev. B},
  volume = {109},
  issue = {21},
  pages = {214308},
  numpages = {32},
  year = {2024},
  month = {Jun},
  publisher = {American Physical Society},
  doi = {10.1103/PhysRevB.109.214308},
  url = {https://link.aps.org/doi/10.1103/PhysRevB.109.214308}
}

@article{gambetta_2011,
  title = {Analytic control methods for high-fidelity unitary operations in a weakly nonlinear oscillator},
  author = {Gambetta, J. M. and Motzoi, F. and Merkel, S. T. and Wilhelm, F. K.},
  journal = {Phys. Rev. A},
  volume = {83},
  issue = {1},
  pages = {012308},
  numpages = {13},
  year = {2011},
  month = {Jan},
  publisher = {American Physical Society},
  doi = {10.1103/PhysRevA.83.012308},
  url = {https://link.aps.org/doi/10.1103/PhysRevA.83.012308}
}

@article{chen_2016,
  title = {Measuring and Suppressing Quantum State Leakage in a Superconducting Qubit},
  author = {Chen, Zijun and Kelly, Julian and Quintana, Chris and Barends, R. and Campbell, B. and Chen, Yu and Chiaro, B. and Dunsworth, A. and Fowler, A. G. and Lucero, E. and Jeffrey, E. and Megrant, A. and Mutus, J. and Neeley, M. and Neill, C. and O'Malley, P. J. J. and Roushan, P. and Sank, D. and Vainsencher, A. and Wenner, J. and White, T. C. and Korotkov, A. N. and Martinis, John M.},
  journal = {Phys. Rev. Lett.},
  volume = {116},
  issue = {2},
  pages = {020501},
  numpages = {5},
  year = {2016},
  month = {Jan},
  publisher = {American Physical Society},
  doi = {10.1103/PhysRevLett.116.020501},
  url = {https://link.aps.org/doi/10.1103/PhysRevLett.116.020501}
}

@article{rol_2019,
  title = {Fast, High-Fidelity Conditional-Phase Gate Exploiting Leakage Interference in Weakly Anharmonic Superconducting Qubits},
  author = {Rol, M. A. and Battistel, F. and Malinowski, F. K. and Bultink, C. C. and Tarasinski, B. M. and Vollmer, R. and Haider, N. and Muthusubramanian, N. and Bruno, A. and Terhal, B. M. and DiCarlo, L.},
  journal = {Phys. Rev. Lett.},
  volume = {123},
  issue = {12},
  pages = {120502},
  numpages = {6},
  year = {2019},
  month = {Sep},
  publisher = {American Physical Society},
  doi = {10.1103/PhysRevLett.123.120502},
  url = {https://link.aps.org/doi/10.1103/PhysRevLett.123.120502}
}

@article{mundada_2019,
  title = {Suppression of Qubit Crosstalk in a Tunable Coupling Superconducting Circuit},
  author = {Mundada, Pranav and Zhang, Gengyan and Hazard, Thomas and Houck, Andrew},
  journal = {Phys. Rev. Appl.},
  volume = {12},
  issue = {5},
  pages = {054023},
  numpages = {10},
  year = {2019},
  month = {Nov},
  publisher = {American Physical Society},
  doi = {10.1103/PhysRevApplied.12.054023},
  url = {https://link.aps.org/doi/10.1103/PhysRevApplied.12.054023}
}

@article{yang_coupler-assisted_2024,
	title = {Coupler-{Assisted} {Leakage} {Reduction} for {Scalable} {Quantum} {Error} {Correction} with {Superconducting} {Qubits}},
	volume = {133},
	url = {https://link.aps.org/doi/10.1103/PhysRevLett.133.170601},
	doi = {10.1103/PhysRevLett.133.170601},
	number = {17},
	urldate = {2026-06-26},
	journal = {Phys. Rev. Lett.},
	publisher = {American Physical Society},
        author = {Yang, Xiaohan and Chu, Ji and Guo, Zechen and Huang, Wenhui and Liang, Yongqi and Liu, Jiawei and Qiu, Jiawei and Sun, Xuandong and Tao, Ziyu and Zhang, Jiawei and others},
	month = oct,
	year = {2024},
	pages = {170601},
}

@article{lacroix2025,
  title = {Fast Flux-Activated Leakage Reduction for Superconducting Quantum Circuits},
  author = {Lacroix, Nathan and Hofele, Luca and Remm, Ants and Benhayoune-Khadraoui, Othmane and McDonald, Alexander and Shillito, Ross and Lazar, Stefania and Hellings, Christoph and Swiadek, Francois  and Colao-Zanuz, Dante and others}, 
  journal = {Phys. Rev. Lett.},
  volume = {134},
  issue = {12},
  pages = {120601},
  year = {2025},
  month = {Mar},
  publisher = {American Physical Society},
  doi = {10.1103/PhysRevLett.134.120601},
}

@article{xin2025,
      title={Improved error correction with leakage reduction units built into qubit measurement in a superconducting quantum processor}, 
      author={Yuejie Xin and Sean L. M. van der Meer and Marc Serra-Peralta and Tim H. F. Vroomans and Matvey Finkel and Hendrik M. Veen and Mark W. Beekman and Leonardo DiCarlo},
      year={2025},
      journal={arXiv:2511.17460},
      url={https://arxiv.org/abs/2511.17460}, 
      pages = {},
      volume = {},
}

@article{chen_fast_2024,
	title = {Fast unconditional reset and leakage reduction in fixed-frequency transmon qubits},
	doi = {10.48550/arXiv.2409.16748},
	urldate = {2026-06-27},
	journal = {arXiv:2409.16748},
	author={Liangyu Chen and Simon Pettersson Fors and Zixian Yan and Anaida Ali and Tahereh Abad and Amr Osman and Eleftherios Moschandreou and Benjamin Lienhard and Sandoko Kosen and Hang-Xi Li and others},
	month = oct,
	year = {2024},
        pages = {},
        volume= {},
}

@article{thorbeck24,
	title = {High-fidelity gates in a transmon using bath engineering for passive leakage reset},
	doi = {10.48550/arXiv.2411.04101},
	author = {Thorbeck, Ted and McDonald, Alexander and Lanes, O. and Blair, John and Keefe, George and Stabile, Adam A. and Royer, Baptiste and Govia, Luke C. G. and Blais, Alexandre},
	month = nov,
	year = {2024},
	journal = {arXiv:2411.04101},
        pages = {}
	
}

@article{bultink2020,
author = {C. C. Bultink  and others},
title = {Protecting quantum entanglement from leakage and qubit errors via repetitive parity measurements},
journal = {Sci. Adv.},
volume = {6},
number = {12},
pages = {eaay3050},
year = {2020},
doi = {10.1126/sciadv.aay3050}}

@Article{McEwen2021,
author={McEwen, M. and others},
title={Removing leakage-induced correlated errors in superconducting quantum error correction},
journal={Nat. Commun.},
year={2021},
month={Mar},
day={19},
volume={12},
number={1},
pages={1761},
issn={2041-1723},
doi={10.1038/s41467-021-21982-y},
url={https://doi.org/10.1038/s41467-021-21982-y}
}

@article{reed_fast_2010,
	title = {Fast reset and suppressing spontaneous emission of a superconducting qubit},
	volume = {96},
	issn = {0003-6951, 1077-3118},
	url = {https://pubs.aip.org/apl/article/96/20/203110/119911/Fast-reset-and-suppressing-spontaneous-emission-of},
	doi = {10.1063/1.3435463},
	number = {20},
	urldate = {2026-06-28},
	journal = {App. Phys. Lett.},
	author = {Reed, M. D. and Johnson, B. R. and Houck, A. A. and DiCarlo, L. and Chow, J. M. and Schuster, D. I. and Frunzio, L. and Schoelkopf, R. J.},
	month = may,
	year = {2010},
	pages = {203110},
}

@article{QuantumToolbox.jl2025,
  title = {Quantum{T}oolbox.jl: {A}n efficient {J}ulia framework for simulating open quantum systems},
  author = {Mercurio, Alberto and Huang, Yi-Te and Cai, Li-Xun and Chen, Yueh-Nan and Savona, Vincenzo and Nori, Franco},
  journal = {{Quantum}},
  issn = {2521-327X},
  publisher = {{Verein zur F{\"{o}}rderung des Open Access Publizierens in den Quantenwissenschaften}},
  volume = {9},
  pages = {1866},
  month = sep,
  year = {2025},
  doi = {10.22331/q-2025-09-29-1866},
  url = {https://doi.org/10.22331/q-2025-09-29-1866}
}

@article{zhou_rapid_2021,
	title = {Rapid and unconditional parametric reset protocol for tunable superconducting qubits},
	volume = {12},
	copyright = {2021 The Author(s)},
	issn = {2041-1723},
	url = {https://www.nature.com/articles/s41467-021-26205-y},
	doi = {10.1038/s41467-021-26205-y},
	number = {1},
	urldate = {2026-06-28},
	journal = {Nat. Commun.},
	publisher = {Nature Publishing Group},
	author = {Zhou, Yu and Zhang, Zhenxing and Yin, Zelong and Huai, Sainan and Gu, Xiu and Xu, Xiong and Allcock, Jonathan and Liu, Fuming and others},
	month = oct,
	year = {2021},
	pages = {5924},
}

\end{document}